\documentclass[preprint,12pt]{elsarticle}
\usepackage{amsmath}
\usepackage{graphicx}
\usepackage{dcolumn}
\usepackage{bm}
\usepackage{caption}
\usepackage{float}
\usepackage{adjustbox}
\usepackage{xcolor}
 \RequirePackage{mathptmx} 

\begin{document}
\begin{frontmatter}
\title{Stellar Weak Rates and Mass Fractions of 20 Most Important \textit{fp}-shell Nuclei with \textit{A} $<$ 65}
\author{Asim Ullah$^{1}$ and Jameel-Un Nabi$^{2}$ }

\address{$^1$ Faculty of Engineering Sciences, Ghulam Ishaq Khan Institute of Engineering Sciences and Technology,
	Topi, 23640, KP,
	Pakistan.}
\address{$^2$University of Wah, Quaid Avenue, Wah Cantt 47040, Punjab, Pakistan}

\begin{abstract}
This work presents stellar weak rates and mass fractions of 20 most important electron capture (\textit{ec}) and beta-decay (\textit{bd}) nuclei with \textit{A} $<$ 65, according to a recent study, during the presupernova evolution of massive stars. The mass fractions of these nuclei were calculated using the Saha's equation, which assumes nuclear statistical equilibrium for a set of initial conditions ($T_9$, $\rho$ and $Y_e$) that represents the trajectory which a massive star's central region takes after its silicon core burns. Our computed mass fractions were found in decent comparison in most cases, and up to a factor 4 difference was noted when compared with the Independent Particle Model results.  The weak interaction (\textit{ec} and \textit{bd}) rates were calculated in a \textit{totally} microscopic fashion using the proton-neutron quasiparticle random phase approximation model and without assuming the Brink-Axel hypothesis. The rates were computed for a wide range of density ($10 -10^{11}$) $g/cm^3$ and temperature ($0.01 - 30$) \textit{GK}. In comparison with large-scale shell-model, our computed rates were found bigger at high values of core temperature. The current study may contribute in a more realistic simulation of stellar evolution processes and modeling of core-collapse supernovae.

\end{abstract}
\end{frontmatter}
\section{Introduction}
\label{intro}
\label{intro}
Core-collapse supernovae are
considered to be one of the main contributors to the
formation of heavier elements in the stellar matter. However, core-collapse simulators, to date, find it challenging to successfully transform the collapse into an explosion. The details of the micro-physics in the stellar environment are poorly understood. There are many
complexities involved. Researchers, world-wide, continue their quest for a better understanding of the dynamics of core-collapse. Since the final outcome of the explosion is sensitive to various physical inputs at the start of each stage of the entire process (i.e., collapse, shock formation, and shock propagation), therefore, the computation of the presupernova stellar structure with the best possible physical data currently available is highly desirable. The structure of the presupernova star is altered both by the change in lepton-to-baryon fraction ($Y_e$) and entropy in its interior. Both, smaller core mass and lower entropy of the precollapse iron should favor an explosion. Because of the small size of iron core, the energy loss by the shock in photodisintegrating iron nuclei in the overlying onion-like matter would be correspondingly smaller \cite{Bur83}. As a result, less energy is stored in nuclear excited states in the lower entropy environment of the presupernova as well as the collapsing core. Thus the collapse can proceed to a higher density and can produce a stronger bounce and a more energetic shock wave \cite{Beth79}. In addition, the smaller entropy environment reduces the abundance of free protons (the main sink of electrons via electron capture reactions) and assists in achieving a higher $Y_e$ value at the time of bounce. The entropy profile determines the extent of the convective burning shells and has crucial effects on the presupernova core structure and nucleosynthesis. \\
$\beta$-decay (\textit{bd}) and electron capture (\textit{ec}) rates are
amongst the key nuclear-physics inputs that determine both the $Y_e$ and the entropy at the presupernova phase \cite{Ar96}. They are believed to have several vital effects in the later phases of stellar evolution of the massive stars. The \textit{ec} effectively reduces the number of electrons available for pressure support, while \textit{bd} has a converse action. Both these processes have a direct influence on the overall $Y_e$ ratio of the core of the star. The neutrinos and antineutrinos, produced during these nuclear weak reactions, escape from the star (during the early phases of presupernova evolution) and thereby channel out the entropy and energy away from the core. These weak interaction rates are important not only in the accurate determination of the structure of the stellar core but also play a vital role in the elemental abundance calculations and nucleosynthesis. Weak interactions in presupernova stars are known to be dominated by allowed Fermi and Gamow-Teller (GT) transitions \cite{Beth79}. \\
In the past, several attempts were made to calculate weak interaction rates in stellar environments in order to gain a better knowledge of the stellar dynamics. Fuller, Fowler, and Newman (FFN)
\cite{Ful80,Ful82a,Ful82b,Ful85} performed the pioneering calculation by tabulating the weak interaction rates for a total of 226 nuclei with masses in the range 60 $\ge$ A $\ge$ 21, employing the independent-particle model (IPM) with the help of the then available experimental data for astrophysical applications. Their rates resulted in a sizable decrease in the $Y_e$ fraction throughout the stellar core, contributing to a refined understanding of the evolution of stars prior to supernova \cite{Wea85}. In 1994, Aufderheide and collaborators \cite{Auf94} investigated the role of weak interaction rates in the evolution of heavy mass stars after silicon burning and compiled a list of most important nuclei (\textit{bd} and \textit{ec}) in $Y_e$ range of 0.40--0.50 employing the IPM model. They extended the FFN work to include heavy nuclei (A $>$ 60) and introduced an  explicit quenching of the GT strength. However, few researchers e.g. \cite{Vet89,El-Kat94,Wil95} highlighted a flaw in the systematic parameterization used by FFN and Aufderheide et al. Later, the proton-neutron quasiparticle random phase approximation (pn-QRPA) \cite{Nab19,Nab04} and shell model \cite{Lan00} computed stellar weak rates and revealed that the GT centroids were misplaced by FFN in some key nuclei which results in disagreement with experimental results. Heger et. al. \cite{Heg01} also conducted simulation studies during the presupernova evolution by estimating weak rates in the mass range 65 $\ge$ A $\ge$ 45 employing large scale shell model \cite{Lan01}. More recently, Nabi and collaborators \cite{Nab21} performed simulation study of the presupernova evolution employing the pn-QRPA computed rates. The pn-QRPA model was used to calculate the stellar weak rates and mass abundances of a total 728 nuclei with mass in the range A = 1–100 and the authors  compiled a list of 50 most important \textit{ec} and \textit{bd} nuclei that had the greatest effect on $Y_e$. The authors compared their calculations with the results of Gross Theory \cite{Fer14} and IPM \cite{Auf94}. The study's conclusion mandates the application of a microscopic model for accurate and reliable computations of stellar weak interaction rates for nuclei possessing astrophysical importance. 

The large-scale shell model (LSSM) and the pn-QRPA are two of the most successful and extensively used models for the microscopic calculation of stellar weak rates.  In this project we focus on the top 50 most important \textit{ec} and \textit{bd} nuclei averaged across the whole stellar route for $0.40 < {Y_e} < 0.50$ (see Table 7 of \cite{Nab21}). In the first phase of our study, we selected 20 most important \textit{fp}-shell nuclei with \textit{A} $<$ 65, that are \textit{ec} and \textit{bd} unstable,  from  the list of nuclei compiled by \cite{Nab21} and present a refined calculation of pn-QRPA rates incorporating latest experimental data and optimum values of model parameters.  Another reason for selecting nuclei with \textit{A} $<$ 65 for the current project was to compare our new calculation with the earlier reported LSSM rates which were performed for the mass range 45 $\le \textit{A} \le $ 65 \cite{Lan00}.

The pn-QRPA model is well suited for stellar rate calculations since it does not use the Brink-Axel hypothesis \cite{Br58} to compute excited state GT strength distributions, as is the case with conventional stellar rate calculations. In order to justify the use of the nuclear model for such a study, earlier a good comparison between the pn-QRPA model  and experimental data was achieved for almost a thousand nuclei (see Figures 11–17 and Tables E–M of \cite{Nab04}). The pn-QRPA model can make use of a large model space (up to 7$\hbar$$\omega$) which permits the calculation of excited states GT strength distribution functions for heavier nuclei. It may be noted that the temperature effects in the current pn-QRPA model appear via a Fermi-Dirac distribution function incorporated in the phase space factors. The pn-QRPA equations do not depend on temperature and allows us to adjust the Fermi and Gamow-Teller amplitudes to experimental values under terrestrial conditions. Calculations using a finite-temperature version of proton-neutron QRPA show that GT strength functions can evolve as the pairing force weakens with temperature (see e.g. \cite{Dzh20}). Such calculations showed that increasing temperature shifts the GT resonance to lower energies and made room for low- and negative-energy GT transitions. However, finite-temperature QRPA calculations have their own shortcomings and it was later commented that correlations beyond thermal QRPA led to significantly higher electron capture rates under the typical thermodynamical conditions \cite{Lit21}. The mass fractions were calculated using Saha's equation under the assumption of nuclear statistical equilibrium. The mass fractions are sensitive to nuclear partition functions, which in turn are heavily dependent on the input mass models \cite{Rau00,Rau03}. In the current study, all theoretical masses (including Q-values) were taken from the recent mass compilation of Audi et al. \cite{Aud17}. In case any mass values were not reported in Audi et al., they were adopted from Ref. \cite{Mol16} to compute the Q-value of the decay reactions.

The paper is structured as follows. Section~2 provides a brief overview of the formalism used in our calculation. Section~3 discusses and compares our findings with previous calculations. The final section contains summary and concluding remarks.

\section{Formalism}
\label{sec:1}
The QRPA computations include two main steps. The quasi-particle (q.p) states were specified in terms of nucleons states using Bogoliubov transformation and then RPA equation was solved in the basis of proton-neutron q.p pairs. The current pn-QRPA model uses separable forces (particle-hole (\textit{ph}) and particle-particle (\textit{pp}) channels) to compute the strength functions of the GT transitions. The resulting RPA matrix equation is thereby transformed into an algebraic equation which is easily solvable.
We began with a spherical nucleon basis ($c^{\dagger}_{jm}$, $c_{jm}$), with $j$ as total
angular momentum and $z$-component $m$. 
The spherical basis was transformed to the (axial-symmetric) deformed basis, denoted by ($d^{\dagger}_{m\alpha}$, $d_{m\alpha}$), using the transformation equation
\begin{equation}\label{df}
	d^{\dagger}_{m\alpha}=\Sigma_{j}D^{m\alpha}_{j}c^{\dagger}_{jm},
\end{equation}
where $d^{\dagger}$ and $c^{\dagger}$ are particle creation operators in the deformed and spherical basis, respectively. The matrices $D^{m\alpha}_{j}$ were obtained by diagonalizing the Nilsson Hamiltonian. The BCS
computation for the proton and neutron systems was performed separately. We took a constant pairing force of strength G ($G_p$ and $G_n$ for protons and neutrons, respectively),
\begin{eqnarray}\label{pr}
	V_{pair}=-G\sum_{jmj^{'}m^{'}}(-1)^{l+j-m}c^{\dagger}_{jm}c^{\dagger}_{j-m}\\ \nonumber
	(-1)^{l^{'}+j^{'}-m^{'}} c_{j^{'}-m^{'}}c_{j^{'}m^{'}},
\end{eqnarray}
the summation over $m$ and $m^{'}$ was limited to $m$, $m^{'}$ $>$ 0, and $l$ is the orbital angular momentum. A q.p basis $(a^{\dagger}_{m\alpha}, a_{m\alpha})$ was later introduced from the Bogoliubov transformation
\begin{equation}\label{qbas}
	a^{\dagger}_{m\alpha}=u_{m\alpha}d^{\dagger}_{m\alpha}-v_{m\alpha}d_{\bar{m}\alpha}
\end{equation}
\begin{equation}
	a^{\dagger}_{\bar{m}\alpha}=u_{m\alpha}d^{\dagger}_{\bar{m}\alpha}+v_{m\alpha}d_{m\alpha},
\end{equation}
where $\bar{m}$, $a^{\dagger}$ and $a$ represent the time reversed state of $m$, the q.p. creation and annihilation operator, respectively which comes in the RPA equation. The occupation amplitudes ($u_{m\alpha}$ and $v_{m\alpha}$) were computed using BCS approximation (satisfying $u^{2}_{m\alpha}$+$v^{2}_{m\alpha}$ = 1).\\
Within the pn-QRPA framework, the GT transitions are described in terms of phonon creation and one describes the QRPA phonons as
\begin{equation}\label{co}
	A^{\dagger}_{\omega}(\mu)=\sum_{pn}[X^{pn}_{\omega}(\mu)a^{\dagger}_{p}a^{\dagger}_{\overline{n}}-Y^{pn}_{\omega}(\mu)a_{n}a_{\overline{p}}],
\end{equation} 
where the indices $n$ and $p$ stand for $m_{n}\alpha_{n}$ and $m_{p}\alpha_{p}$, respectively, differentiating neutron and proton single-particle states. The summation was taken over all proton-neutron pairs satisfying $\mu=m_{p}-m_{n}$ and $\pi_{p}.\pi_{n}$=1, with $\pi$ representing parity. In Eq.~(\ref{co}), $X$ and $Y$ represent the forward- and backward-going amplitudes, respectively, and are the eigenfunctions of the RPA matrix equation, given as:
\begin{equation}\label{ME}
	\begin{pmatrix}
		A & B \\ -B & -A
	\end{pmatrix}
	\begin{pmatrix}
		X \\ Y
	\end{pmatrix}
	=\omega
	\begin{pmatrix}
		X \\ Y
	\end{pmatrix}.
\end{equation}
As mentioned above, in our model the proton-neutron residual interactions work through \textit{ph} and \textit{pp} channels, characterized by interaction constants $\chi$ and $\kappa$, respectively. The $ph$ GT force can be expressed as
\begin{equation}\label{ph}
	V^{ph}= +2\chi\sum^{1}_{\mu= -1}(-1)^{\mu}Y_{\mu}Y^{\dagger}_{-\mu},\\
\end{equation}
with
\begin{equation}\label{y}
	Y_{\mu}= \sum_{j_{p}m_{p}j_{n}m_{n}}<j_{p}m_{p}\mid
	t_- ~\sigma_{\mu}\mid
	j_{n}m_{n}>c^{\dagger}_{j_{p}m_{p}}c_{j_{n}m_{n}},
\end{equation}
and the $pp$ GT force as
\begin{equation}\label{pp}
	V^{pp}= -2\kappa\sum^{1}_{\mu=
		-1}(-1)^{\mu}P^{\dagger}_{\mu}P_{-\mu},
\end{equation}
with
\begin{eqnarray}\label{p}
	P^{\dagger}_{\mu}= \sum_{j_{p}m_{p}j_{n}m_{n}}<j_{n}m_{n}\mid
	(t_- \sigma_{\mu})^{\dagger}\mid
	j_{p}m_{p}>\times \nonumber\\
	(-1)^{l_{n}+j_{n}-m_{n}}c^{\dagger}_{j_{p}m_{p}}c^{\dagger}_{j_{n}-m_{n}}.
\end{eqnarray}
Here, the different signs in \textit{ph} and \textit{pp} force reveal the opposite nature of these interactions i.e. \textit{pp} force is attractive while the \textit{ph} force is repulsive. The interaction constants $\chi$ and $\kappa$ were chosen in concordance with the suggestion given in Ref. \cite{JUN}. 
The pn-QRPA equations strongly depend on the  \textit{pp} channel. If  \textit{pp} channel is strongly attractive, the QRPA equations leads to imaginary eigenvalues for the lowest solutions and the ground state correlations are much enhanced, violating the basic assumption of RPA. However, in our formalism, even if the lowest solution enters the non-physical region, the other solution with positive eigenvalues can still be obtained. This means, on the other hand, an occurrence of the RPA collapse is not apparent. This behavior of the lowest-lying solutions can be seen once the  \textit{pp} strength is changed step by step \cite{Mut92}. Further the Ikeda sum rule was satisfied for all the cases considered in our calculation and this happens when all the QRPA solutions are in positive energy region \cite{Mut92}. For separable forces, the matrix equation (Eq. \ref{ME}) can be given explicitly as:
\begin{equation}\label{x}
	\begin{split}
		X^{pn}_{\omega}=\frac{1}{\omega-\varepsilon_{pn}}[2\chi(q_{pn}Z^{-}_{\omega}+\tilde{q_{pn}}Z^{+}_\omega)\\
		-2\kappa(q^{U}_{pn}Z^{- -}_{\omega}+q^{V}_{pn}Z^{+ +}_{\omega})],
	\end{split}
\end{equation}
\begin{equation}\label{y}
	\begin{split}
		Y^{pn}_{\omega}=\frac{1}{\omega+\varepsilon_{pn}}[2\chi(q_{pn}Z^{+}_{\omega}+\tilde{q_{pn}}Z^{-}_\omega)\\
		+2\kappa(q^{U}_{pn}Z^{+ +}_{\omega}+q^{V}_{pn}Z^{- -}_{\omega})],
	\end{split}
\end{equation}
where $\varepsilon_{pn}=\varepsilon_{p}+\varepsilon_{n}$,\\ \\
$q_{pn}=f_{pn}u_pv_n$, $q_{pn}^{U}=f_{pn}u_pu_n$,\\ \\
$\tilde q_{pn}=f_{pn}v_pu_n$, $q_{pn}^{V}=f_{pn}v_pv_n$,\\ \\
\begin{equation}\label{Z-}
	Z^{-}_{\omega}= \sum_{pn}(X^{pn}_{\omega}q_{pn}-Y^{pn}_{\omega}\tilde q_{pn})\\
\end{equation}
\begin{equation}\label{Z+}
	Z^{+}_{\omega}= \sum_{pn}(X^{pn}_{\omega}\tilde q_{pn}-Y^{pn}_{\omega}q_{pn})\\
\end{equation}
\begin{equation}\label{Z--}
	Z^{- -}_{\omega}= \sum_{pn}(X^{pn}_{\omega}q^{U}_{pn}+Y^{pn}_{\omega}q^{V}_{pn})\\
\end{equation}
\begin{equation}\label{Z++}
	Z^{+ +}_{\omega}=
	\sum_{pn}(X^{pn}_{\omega}q^{V}_{pn}+Y^{pn}_{\omega}q^{U}_{pn}).
\end{equation}
We finally get a matrix equation,
\begin{equation}\label{M}
	M z=0,
\end{equation}
with solution given by
\begin{equation}\label{DM0}
	|M|=0.
\end{equation}
Consequently, the RPA matrix equation (Eq.~\ref{ME}) is reduced to determining roots of algebraic equation (Eq.~\ref{DM0}).
In the absence of \textit{pp} force ($\kappa$ = 0), Eq.~\ref{DM0} is a second order equation while the equation is of fourth-order once the \textit{pp} force is switched on. For more information on the solution of Eq.~\ref{DM0}, we refer to Ref.~\cite{Mut92}. 

The weak decay rate from the $\mathit{m}$th state (parent nucleus) to
the $\mathit{n}$th state (daughter nucleus) in stellar environment can be determined using the formula given below

\begin{eqnarray}\label{rate}
	\lambda^{ec(bd)}_{mn} =\frac{ln2}{D}[{\phi_{mn}^{ec(bd)}(\rho, T, E_{f})}] \nonumber \\ 
	\times [B(F)_{mn}+\frac{B(GT)_{mn}}{(g_{A}/g_{V})^{-2}} ].
\end{eqnarray}

The value of constant D appearing in Eq.~\ref{rate} was taken as 6143 s \cite{Har09}. The value of $g_{A}$/$g_{v}$, representing the ratio of axial and vector coupling constant, was taken as -1.2694 \cite{Nak10}. $B(F)_{mn}$ and $B(GT)_{mn}$ are the total reduced transition probabilities due to Fermi interactions and GT interactions, respectively. 

\begin{equation}
	B(F)_{mn} = [{2J_{m}+1}]^{-1}  \mid<n \parallel \sum_{k}t_{\pm}^{k}
	\parallel m> \mid ^{2}
\end{equation}

\begin{equation}\label{bgt}
	B(GT)_{mn} = [{2J_{m}+1}]^{-1}  \mid <n
	\parallel \sum_{k}t_{\pm}^{k}\vec{\sigma}^{k} \parallel m> \mid ^{2},
\end{equation}
where $J_m$, $t_\pm$ and $\vec{\sigma}$ stands for the total spin of parent nucleus, isospin operators (raising/lowering) and Pauli spin matrices, respectively.\\
The phase space integrals $(\phi_{mn})$ for \textit{ec} and \textit{bd} are; (hereafter natural units are used, $\hbar=c=m_{e}=1$)
\begin{equation}\label{pc}
	\phi^{ec}_{mn} = \int_{w_{l}}^{\infty} w (w_{m}+w)^{2}({w^{2}-1})^{\frac{1}{2}} F(+Z,w) G_{-} dw,
\end{equation}

\begin{equation}\label{ps}
	\phi^{bd}_{mn} = \int_{1}^{w_{m}} w (w_{m}-w)^{2}({w^{2}-1})^{\frac{1}{2}} F(+Z,w)
	(1-G_{-}) dw,
\end{equation}
In Eqs.~\ref{pc} and \ref{ps}, $w$, $w_l$ and $w_{m}$ stands for total energy (kinetic+rest mass) of the electron, total threshold energy for \textit{ec} and the total energy of \textit{bd}, respectively. The Fermi functions ($F(+ Z, w)$) were calculated employing the procedure used by Gove and Martin ~\cite{Gov71}. $G_{-}$ represents the electron distribution function expressed as
\begin{equation}
	G_- = [1 + exp(\frac{E-E_f}{kT})]^{-1},
\end{equation}
where $E~(= w-1)$ and $E_f$ are the kinetic and Fermi energy of electrons, respectively.

%
The \textit{ec} and \textit{bd} total weak-interaction rates were calculated using
\begin{equation}\label{ecbd}
	\lambda^{ec(bd)} =\sum _{mn}P_{m} \lambda^{ec(bd)} _{mn}.
\end{equation}
The summation was applied over all the initial and final states and satisfactory convergence in \textit{ec} and \textit{bd} rates was achieved. $P_{m} $ in the above equation denotes the occupation probability of parent excited states and follows the normal Boltzmann distribution.
%

The mass fractions of $m^\textit{th}$ nucleus can be calculated using the Saha's equation expressed as 
\begin{multline}\label{x1}
	x_m(A,Z)= \frac {G_m(A,Z,T)}{2}\left(\frac{\rho N_{A} \lambda_T^{3}}{2}\right)^{A-1} \\
	\times A^{\frac{5}{2}}{x_{n}^{A-Z}x_{p}^{Z}}\exp
	\left[ Q_m(A,Z)/k_BT \right],
\end{multline}
where $G_m$(A,Z,T) denotes the nuclear partition function (NPF) of the $m^\textit{th}$ nucleus which was computed using a newly recipe introduced by nabi and collaborators \cite{Nab16a,Nab16b}. $Q_m$ stands for the ground state binding energy of the nucleus \textit{m}. $N_A$ is the Avogadro's number while $\lambda_T$ $(= [{2{h^{-2}}{\pi m_Hk_BT}}]^{\frac{-1}{2}})$ denotes the thermal wavelength. The mass fractions of free proton ($x_p$) and free neutron ($x_n$) can be computed subject to mass and charge conservation, respectively given by 
\begin{equation}
	\sum_{m} x_m = 1,
\end{equation} 

\begin{equation}
	\sum_{m} \frac{Z_m}{A_m}x_m = Y_e = \frac{1-\eta}{2},
\end{equation}
where $\eta$ stands for neuron excess.  Mass fractions of any nucleus can be computed after $x_n$ and $x_p$ are determined. 
We refer to~\cite{Nab21} for further details of the underlying formalism.
\section{Results and Discussion}
\begin{figure*}[]
		\centering
		\includegraphics[width=0.80\textwidth]{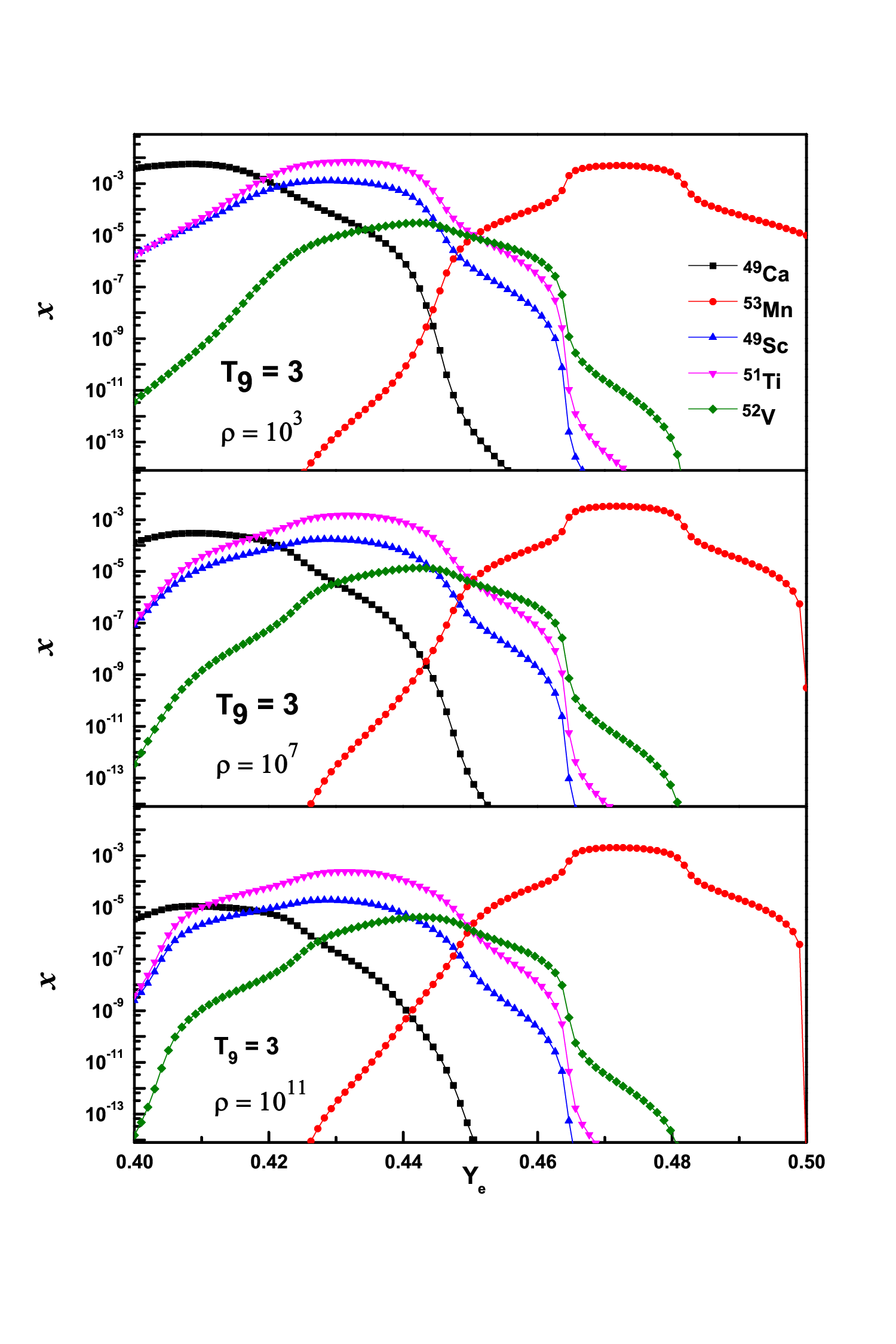}
		\caption{Calculated mass fractions for selected nuclei, as a function of $Y_e$ and $\rho$, at fixed core temperature. The units of $\rho$ and $T_9$ are $g/cm^3$ and \textit{GK}, respectively. The mass fractions are shown in log scale. \label{f1_log} }
	\end{figure*}

\begin{figure*}[]
		\centering
		\includegraphics[width=0.80\textwidth]{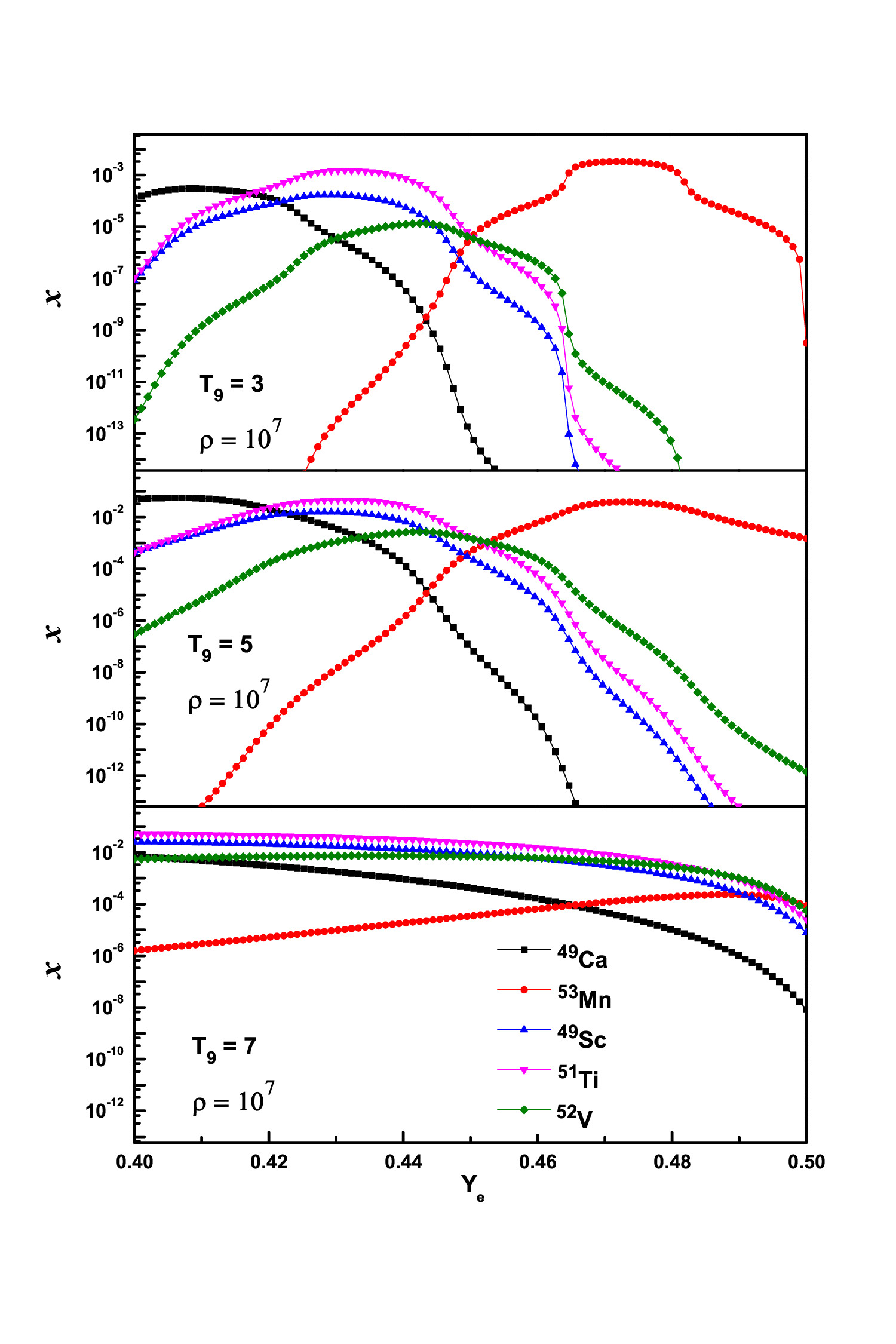}
		\caption{Calculated mass fractions for selected nuclei, as a function of $Y_e$ and $T_9$, at fixed stellar density. The units of $T_9$ and $\rho$ are \textit{GK} and $g/cm^3$, respectively. The mass fractions are shown in log scale. \label{f2_log} }
	\end{figure*}

We present the computation of stellar weak rates and mass fractions of the 20 most important \textit{ec} and \textit{bd} \textit{fp}-shell nuclei with \textit{A} $<$ 65. These nuclei were selected from the published  top 50 list (see Table~7 of \cite{Nab21}) and are reproduced in Table~\ref{T1}. The rank of nuclei mentioned in Table~\ref{T1} was adopted from \cite{Nab21}. The stellar weak rates (\textit{ec} and \textit{bd}) were computed employing the pn-QRPA model and the results were compared with those of large-scale shell model \cite{Lan00} (referred to as LSSM throughout the text). The mass fractions were calculated employing the Saha’s equation under the assumption of nuclear statistical equilibrium and were compared with the IPM results \cite{Auf94}, referred to as IPM here onwards. \\
Fig. \ref{f1_log} displays our calculated mass fractions for most abundant nuclei as a function of $Y_e$ at different densities ($\rho$ = $10^{3}$, $10^{7}$ \& $10^{11}$) $g/cm^3$ and fixed temperature ($T_9$ = 3) \textit{GK}. Fig. \ref{f2_log} shows mass fractions for the same nuclei as a function of $Y_e$ at different temperatures ($T_9$ = 3, 5 \& 7) \textit{GK} and fixed density ($\rho$ = $10^{7}$) $g/cm^3$. It can be seen from these figures that there is slight increase in mass fractions with increase in $T_9$ values. This behavior is in accordance with the findings of Ref.~\cite{Sei09}. As the core density increases, the mass fractions decrease considerably. This is because with increase in core density, re-distribution of the mass occurs and new nuclei begin to appear because of the assumption of  nuclear statistical equilibrium. This behavior can also be seen in the IPM calculation. 

Tables~(\ref{T1MFec}-\ref{T2MFbd}) depict the comparison of our computed mass fractions with those of IPM results (wherever available). Eight different sets of initial conditions including $\rho$, $T_9$ and $Y_e$ were selected  to compute the mass fractions by IPM. In order to compare the two calculations, we chose the same initial conditions. These sets mainly describe the trajectory that a massive star's central region follows after its silicon core burns. Table~\ref{T1MFec}~\&~\ref{T2MFec} compare mass fraction results of the most important \textit{ec} nuclei, whereas Table~\ref{T1MFbd}~\&~\ref{T2MFbd} compare those of \textit{bd} nuclei. It can be seen from the tables that our computed mass fractions are, at times a factor 4 smaller and at times comparable to the IPM results. The use of different recipes for computation of  nuclear partition function (no discrete excited state was calculated by IPM and an integration was performed on all excited states using a back-shifted Fermi gas level density formula) and usage of different mass models could be the reason for the discrepancy in the two computations. The pool of nuclei considered by \cite{Nab21} was 728 and that by IPM was 150. \\
We compare our computed \textit{ec} and \textit{bd} rates for the selected \textit{fp}-shell nuclei with those of LSSM rates in Tables~(\ref{T1ec}-\ref{T2bd}). It is worth mentioning that the same set of physical conditions  were used for comparing the two microscopic calculations. Tables~\ref{T1ec}~\&~\ref{T2ec} show the comparison of \textit{ec} rates for the 20 important \textit{ec} nuclei at $T_9$ (= 1, 3, 10 \& 30) \textit{GK} and $\rho$$Y_e$ range ($\rho$$Y_e$ = $10^{7}-10^{10}$) $g/cm^3$, whereas Tables~\ref{T1bd}~\&~\ref{T2bd} show the comparison of \textit{bd} rates for the important \textit{bd} nuclei at the same $T_9$ and $\rho$$Y_e$ values. These tables show that as the stellar temperature rises, both the \textit{ec} and \textit{bd} rates on the selected nuclei increase. This is due to the fact that as temperature rises, the occupation probability of the parent excited states increases, contributing effectively to overall rates. The \textit{ec} rates further  increase with rise in $\rho$$Y_e$ values due to increase in electron chemical potential. There is a reduction in the \textit{bd} rates for high $\rho$$Y_e$ values due to decrease in available phase space because of increased electron chemical potential. For the sake of analysis of comparing \textit{ec} rates, we selected one odd-odd ($^{60}$Co) and one even-even ($^{56}$Fe) nucleus. The $^{60}$Co is rated as the top-most important \textit{ec} nucleus according to the survey of IPM while it is 4\textit{th} according to the compilation of Ref.~\cite{Nab21}. For $^{60}$Co at low densities ($\rho$$Y_e$ = $10^{7}$ \& $10^{8}$) $g/cm^3$, the LSSM computed \textit{ec} rates are orders of magnitude bigger than the pn-QRPA rates for low $T_9$ values (1 \& 3) \textit{GK}. However it is to be noted that the computed rates are very small in magnitude, in the two models, and can hardly affect the simulation results. At $T_9$ = 10 \textit{GK} the two computations match well while the pn-QRPA rates are an order of magnitude bigger at $T_9$ = 30 \textit{GK}. For higher densities ($\rho$$Y_e$ = $10^{9}$ \& $10^{10}$) $g/cm^3$, both computed rates are comparable at low $T_9$ values whereas the pn-QRPA rates are an order bigger in magnitude at $T_9$ = 30 \textit{GK}. The $ec$ rates of $^{56}$Fe are of considerable astrophysical importance in the presupernova evolution of a massive star. It is rated 12\textit{th} in the IPM while 21\textit{st} according to compilation of Ref.~\cite{Nab21}. For this case, at all $\rho$$Y_e$ values, the pn-QRPA rates are up to an order of magnitude smaller than LSSM rates at $T_9$ = 1 \textit{GK}. It is again to be to be noted that this difference is not significant keeping in view that the magnitude of the computed rates is too small. At $T_9$ = 3 \& 10 \textit{GK}, the two computations are in good agreement. An order of magnitude bigger pnQRPA rates are noted at $T_9$ = 30 \textit{GK}.

For presenting a comparison of the calculated \textit{bd} rates in the two models, we selected one even-odd  ($^{57}$Fe) and one odd-even ($^{59}$Mn) nucleus. Both these cases are among the top 20 important \textit{bd} nuclei according to both IPM and Ref.~\cite{Nab21} compilations. Considering the case of $^{57}$Fe, for all $\rho$$Y_e$ values, the pn-QRPA computed \textit{bd} rates are  smaller at $T_9$ = 1, 3 \& 10 \textit{GK}. At $T_9$ = 30 \textit{GK}, the pn-QRPA rates are up to an order of magnitude bigger than LSSM \textit{bd} rates. For $^{59}$Mn, the \textit{bd} rates of pn-QRPA are comparable to LSSM rates at temperature ($T_9$ = 1, 3, \& 10) \textit{GK} at almost all $\rho$$Y_e$ values. For high temperature ($T_9$ = 30) \textit{GK} the pn-QRPA computed rates are an order of magnitude bigger than the LSSM rates at all $\rho$$Y_e$ values.

To summarize, the pn-QRPA computed \textit{ec} rates for almost all the cases shown in Tables~\ref{T1ec}-\ref{T2ec} are bigger than the corresponding LSSM rates at all $T_9$ and $\rho$$Y_e$ values (except $^{56}$Fe, $^{60}$Co, $^{56}$Ni, $^{52}$V, $^{50}$Sc and $^{56}$Mn). The enhancement factor is high for high $\rho$$Y_e$ values. The pn-QRPA model calculates small $ec$ rates at low temperatures compared to LSSM rates. These are however small numbers and hardly can affect the simulation results. Our computed \textit{bd} rates for most of the cases shown in Tables~\ref{T1bd}-\ref{T2bd} are bigger than LSSM rates at all $T_9$ and $\rho$$Y_e$ values except at $\rho$$Y_e$ = 10$^{10}$ $g/cm^3$. Again few exceptional cases ($^{53}$Mn, $^{64}$Co, $^{51}$Sc, $^{57}$Fe and $^{57}$Co) were there, for which the LSSM rates were found bigger at low core temperatures. At higher temperature ($T_9$ $>$ 5) \textit{GK}, excited states do play significant role in convergence of rates. The LSSM approach applied the Lanczos method to derive GT strength function but was limited up to 100 iterations which were insufficient for converging the states above 2.5 MeV excitation energies. Moreover, LSSM used the Brink-Axel hypothesis \cite{Br58} to estimate contribution from higher excited states at high temperatures and densities. These may be cited as the reasons for the lower rates of LSSM at high $T_9$ and $\rho$$Y_e$ values. The pn-QRPA approach had no such model limitations. Moreover, the pn-QRPA model involves a large model space (up to 7$\hbar \omega$) to adequately handle excited states in parent and daughter nuclei. We emphasize that Brink-Axel hypothesis was not employed in calculation of excited state GT strength functions in our computation \cite{Nab21+}.

\section{Summary and Conclusion} 
In this study we provide the calculations of stellar weak rates and mass fractions of 20 most important \textit{fp}-shell nuclei of astrophysical significance with \textit{A} $<$ 65. The selected nuclei were the ones having largest effect on the $Y_e$ according to the compilation of Ref.~\cite{Nab21}. The mass fractions were computed from the Saha’s equation under the assumption of nuclear statistical equilibrium and were compared with the IPM calculation. It was found that the calculated mass fractions of both the models were in fair comparison (within a factor 4). Our calculated mass fractions used a novel technique for computing nuclear partition functions. We treated all excited states up to excitation energy of 10 MeV as \textit{discrete} and performed \textit{summation} over the states. This may lead to a more reliable computation of isotopic abundances according to a previous study. \\
The \textit{ec} and \textit{bd} rates, for the selected nuclei, were calculated for a wide range of $\rho$$Y_e$ ($10-10^{11}$) $g/cm^3$ and $T_9$ (0.1 - 30) \textit{GK} and were compared with the  LSSM rates. It was found that our computed \textit{ec} and \textit{bd} rates were most of the time bigger than LSSM rates (up to an order of magnitude) except for few cases at low $T_9$ values. The problem of convergence of GT strength functions and use of Brink-Axel hypothesis in the rate calculation could be the reasons for the smaller LSSM numbers. During the later stages of the stellar evolution, the contribution from excited states becomes significant due to high temperature and can efficiently increase the stellar weak rates. We computed the excited state GT strength functions in a microscopic fashion without incorporating the Brink-Axel hypothesis in our calculation of weak rates. It is hoped that the present study, due to it's increased reliability, could prove useful for simulation of the stellar evolution processes and modeling of  core-collapse supernovae. A similar comparison of weak rates for $A >$ 65 with previous calculations is currently in progress and would be reported once completed.


	\newpage
	\begin{table*}[]
		\caption{ The 20 most important \textit{fp}-shell nuclei \cite{Nab21} considered in this project. The electron capture  (\textit{ec}) and $\beta$-decay  (\textit{bd}) nuclei are shown in separate columns.}
		\label{T1}
		\centering
		\begin{tabular}{ccc|ccc}
			
			\hline
			\multicolumn{3}{c}{\textit{ec} nuclei} & \multicolumn{3}{c}{\textit{bd} nuclei}   \\
			
			\hline
			A   & Symbol  & Rank  & A   & Symbol & Rank  \\
			56 & Mn & 1  & 49 & Sc & 2  \\
			52 & V  & 2  & 65 & Co & 3  \\
			60 & Co & 4  & 63 & Co & 4  \\
			53 & Mn & 5  & 50 & Sc & 5  \\
			49 & Sc & 6  & 59 & Mn & 6  \\
			50 & Sc & 8  & 53 & Mn & 7  \\
			55 & Fe & 9  & 64 & Co & 8  \\
			59 & Co & 10 & 49 & Ca & 9  \\
			61 & Ni & 13 & 55 & Mn & 10 \\
			51 & Ti & 16 & 58 & Cr & 11 \\
			56 & Ni & 18 & 61 & Fe & 13 \\
			56 & Fe & 21 & 51 & Ti & 14 \\
			65 & Cu & 22 & 51 & Sc & 15 \\
			55 & Co & 23 & 57 & Fe & 16 \\
			57 & Ni & 25 & 57 & Cr & 17 \\
			54 & Cr & 27 & 55 & Cr & 18 \\
			64 & Cu & 30 & 62 & Fe & 19 \\
			57 & Co & 31 & 57 & Mn & 20 \\
			63 & Cu & 32 & 57 & Co & 21 \\
			58 & Ni & 33 & 53 & Ti & 22\\
			\hline
		\end{tabular}
	\end{table*}
	
	\begin{table*}[]
	
		\caption{ The computed mass fractions of the \textit{ec} nuclei at different $T_9$, $\rho$ and $Y_e$ values. Comparison is shown with the IPM calculation \cite{Auf94} wherever possible. The units of $T_9$ and $\rho$ are \textit{GK} and $g/cm^3$, respectively. The exponents are shown in parenthesis. }
		\label{T1MFec}
		\centering
		\scalebox{0.7}{
			\begin{tabular}{c|cc|cc|cc|cc}
				
				\hline       
				\multicolumn{3}{l}{} & \multicolumn{4}{l}{Mass Fractions (top 20 \textit{ec} nuclei)} & \multicolumn{1}{l}{}         \\
				\hline
				{}      & \multicolumn{2}{c}{$T_9$ = 7.33}     & \multicolumn{2}{c}{$T_9$ = 5.39}     & \multicolumn{2}{c}{$T_9$ = 4.93}     & \multicolumn{2}{c}{$T_9$ = 4.24}     \\
				{} & \multicolumn{2}{c}{$\rho$ = 4.01(+10)} & \multicolumn{2}{c}{$\rho$ = 2.20(+09)} & \multicolumn{2}{c}{$\rho$ = 1.06(+09)} & \multicolumn{2}{c}{$\rho$ = 3.30(+08)} \\
				{}      & \multicolumn{2}{c}{$Y_e$ = 0.41}     & \multicolumn{2}{c}{$Y_e$ = 0.425}    & \multicolumn{2}{c}{$Y_e$ = 0.43}     & \multicolumn{2}{c}{$Y_e$ = 0.44}     \\ 
				& pn-QRPA       & IPM        & pn-QRPA       & IPM        & pn-QRPA       & IPM        & pn-QRPA       & IPM        \\
				\hline
				\hline
				$^{56}$Mn & 5.23(-05) & ---       & 2.85(-04) & 3.09(-04) & 4.17(-04) & 5.62(-04) & 6.53(-04) & 1.12(-03) \\
				$^{52}$V  & 2.08(-04) & 2.08(-04) & 6.01(-04) & 5.61(-04) & 6.50(-04) & 7.40(-04) & 4.71(-04) & 6.76(-04) \\
				$^{60}$Co & 3.85(-06) & ---       & 2.58(-05) & 4.25(-05) & 4.41(-05) & 9.51(-05) & 1.14(-04) & 3.26(-04) \\
				$^{53}$Mn & 6.45(-10) & ---       & 4.12(-09) & ---       & 9.54(-09) & ---       & 1.21(-07) & 2.66(-07) \\
				$^{49}$Sc & 5.44(-03) & 1.39(-02) & 9.28(-03) & ---       & 6.65(-03) & ---       & 1.18(-03) & ---      \\
				$^{50}$Sc & 8.34(-03) & 1.03(-02) & 2.58(-03) & 4.66(-03) & 8.86(-04) & 1.74(-03) & 2.59(-05) & 5.81(-05) \\
				$^{55}$Fe & 2.92(-10) & ---       & 2.25(-09) & ---       & 5.87(-09) & ---       & 1.05(-07) & 1.94(-07) \\
				$^{59}$Co & 3.37(-07) & ---       & 5.57(-06) & 1.25(-05) & 1.49(-05) & 4.42(-05) & 1.33(-04) & 5.16(-04) \\
				$^{61}$Ni & 4.65(-08) & ---       & 6.74(-07) & ---       & 1.76(-06) & 3.41(-06) & 1.67(-05) & 4.39(-05) \\
				$^{51}$Ti & 8.77(-03) & 8.77(-03) & 2.64(-02) & ---       & 2.43(-02) & 2.91(-02) & 7.60(-03) & 9.99(-03) \\
				$^{56}$Ni & 1.31(-20) & ---       & 4.23(-21) & ---       & 5.72(-21) & ---       & 1.22(-19) & ---      \\
				$^{56}$Fe & 1.35(-07) & ---       & 5.12(-06) & ---       & 2.02(-05) & ---       & 5.05(-04) & ---      \\
				$^{65}$Cu & 5.12(-06) & ---       & 4.69(-05) & 7.94(-05) & 8.64(-05) & 1.84(-04) & 2.08(-04) & 5.52(-04) \\
				$^{55}$Co & 7.67(-17) & ---       & 5.43(-17) & ---       & 8.30(-17) & ---       & 1.33(-15) & ---      \\
				$^{57}$Ni & 1.74(-17) & ---       & 1.20(-17) & ---       & 1.86(-17) & ---       & 3.36(-16) & ---      \\
				$^{54}$Cr & 9.48(-04) & ---       & 2.48(-02) & 1.85(-02) & 5.98(-02) & ---       & 2.14(-01) & 2.02(-01) \\
				$^{64}$Cu & 7.20(-08) & ---       & 3.06(-07) & ---       & 4.81(-07) & 5.41(-07) & 1.30(-06) & 1.97(-06) \\
				$^{57}$Co & 3.27(-11) & ---       & 2.25(-10) & ---       & 5.79(-10) & ---       & 1.10(-08) & ---      \\
				$^{63}$Cu & 2.96(-09) & ---       & 2.38(-08) & ---       & 5.48(-08) & ---       & 4.52(-07) & 1.26(-06) \\
				$^{58}$Ni & 3.07(-14) & ---       & 1.74(-13) & ---       & 4.88(-13) & ---       & 1.82(-11) & --- \\
				\hline
		\end{tabular}}
	\end{table*}
	
	\begin{table*}[]
	
		\caption{ Same as Table \ref{T1MFec} but with different $T_9$, $\rho$ and $Y_e$ values.}
		\label{T2MFec}
		\centering
		\scalebox{0.7}{
			\begin{tabular}{c|cc|cc|cc|cc}
				
				\hline
				\multicolumn{3}{l}{} & \multicolumn{4}{l}{Mass Fractions (top 20 \textit{ec} nuclei)} & \multicolumn{1}{l}{}         \\
				\hline
				{}     & \multicolumn{2}{c}{$T_9$ = 3.8}      & \multicolumn{2}{c}{$T_9$ = 3.65}     & \multicolumn{2}{c}{$T_9$ = 3.4}      & \multicolumn{2}{c}{$T_9$ = 3.26}     \\
				{} & \multicolumn{2}{c}{$\rho$ = 1.45(+08)} & \multicolumn{2}{c}{$\rho$ = 1.07(+08)} & \multicolumn{2}{c}{$\rho$ = 5.86(+07)} & \multicolumn{2}{c}{$\rho$ = 4.32(+07)} \\
				{}      & \multicolumn{2}{c}{$Y_e$ = 0.45}     & \multicolumn{2}{c}{$Y_e$ = 0.455}    & \multicolumn{2}{c}{$Y_e$ = 0.47}     & \multicolumn{2}{c}{$Y_e$ = 0.485}    \\
				& pn-QRPA & IPM        & pn-QRPA       & IPM        & pn-QRPA       & IPM        & pn-QRPA       & IPM        \\
				\hline
				\hline
				$^{56}$Mn & 3.75(-04) & 7.03(-04) & 1.46(-04) & 2.77(-04) & 9.37(-09) & 9.37(-09) & 2.80(-15) & 7.09(-15) \\
				$^{52}$V  & 7.18(-05) & 1.17(-04) & 1.80(-05) & 2.95(-05) & 2.35(-10) & 2.35(-10) & 3.09(-17) & ---      \\
				$^{60}$Co & 2.01(-04) & 6.03(-04) & 1.12(-04) & 3.45(-04) & 3.04(-08) & 3.04(-08) & 1.89(-14) & 7.53(-14) \\
				$^{53}$Mn & 4.34(-05) & 8.08(-05) & 1.82(-04) & 3.74(-04) & 6.39(-03) & 1.30(-02) & 2.18(-04) & 4.86(-04) \\
				$^{49}$Sc & 5.71(-06) & ---       & 4.18(-07) & ---       & 1.36(-14) & ---       & 1.62(-23) & ---      \\
				$^{50}$Sc & 5.91(-09) & ---       & 1.37(-10) & ---       & 2.81(-20) & ---       & 4.55(-31) & ---      \\
				$^{55}$Fe & 7.14(-05) & 1.14(-04) & 3.73(-04) & 6.68(-04) & 2.87(-02) & 5.20(-02) & 1.47(-03) & 2.96(-03) \\
				$^{59}$Co & 3.07(-03) & 1.14(-02) & 4.48(-03) & 1.75(-02) & 1.36(-04) & 1.36(-04) & 4.97(-09) & ---      \\
				$^{61}$Ni & 5.75(-04) & 1.46(-03) & 9.43(-04) & 2.55(-03) & 5.22(-05) & ---       & 2.56(-09) & ---      \\
				$^{51}$Ti & 8.32(-05) & 1.28(-04) & 8.08(-06) & 1.18(-05) & 6.49(-13) & ---       & 1.26(-21) & ---      \\
				$^{56}$Ni & 2.16(-14) & ---       & 7.88(-13) & ---       & 3.24(-05) & 2.72(-05) & 1.56(-01) & 1.37(-01) \\
				$^{56}$Fe & 9.34(-02) & 8.30(-02) & 3.06(-01) & 2.89(-01) & 5.29(-01) & 5.29(-01) & 8.17(-04) & 8.64(-04) \\
				$^{65}$Cu & 1.40(-04) & ---       & 5.20(-05) & ---       & 7.60(-10) & ---       & 2.08(-17) & ---      \\
				$^{55}$Co & 2.72(-11) & ---       & 4.67(-10) & ---       & 1.71(-04) & 5.35(-04) & 1.52(-02) & 5.15(-02) \\
				$^{57}$Ni & 1.08(-11) & ---       & 2.11(-10) & ---       & 1.46(-04) & 1.71(-04) & 1.77(-02) & 2.25(-02) \\
				$^{54}$Cr & 1.69(-01) & 1.63(-01) & 7.73(-02) & 7.30(-02) & 4.54(-06) & 4.54(-06) & 1.40(-12) & ---      \\
				$^{64}$Cu & 4.74(-06) & 7.37(-06) & 3.26(-06) & 5.28(-06) & 2.82(-09) & 2.82(-09) & 3.06(-15) & ---      \\
				$^{57}$Co & 1.10(-05) & 3.84(-05) & 6.44(-05) & 2.53(-04) & 9.01(-03) & 3.55(-02) & 6.12(-04) & 2.69(-03) \\
				$^{63}$Cu & 1.97(-05) & 5.16(-05) & 3.41(-05) & 9.55(-05) & 3.06(-06) & ---       & 1.88(-10) & ---      \\
				$^{58}$Ni & 2.19(-07) & ---       & 3.07(-06) & ---       & 6.08(-02) & 5.44(-02) & 2.59(-01) & 2.50(-01)\\
				\hline
		\end{tabular}}
	\end{table*}
	
	\begin{table*}[]
		\caption{ The computed mass fractions of the \textit{bd} nuclei at different $T_9$, $\rho$ and $Y_e$ values. Comparison is shown with the IPM calculation \cite{Auf94} wherever possible. The units of $T_9$ and $\rho$ are \textit{GK} and $g/cm^3$, respectively. The exponents are shown in parenthesis. }
		\label{T1MFbd}
		\centering
		\scalebox{0.8}{
			\begin{tabular}{c|cc|cc|cc|cc}
				
				\hline       
				\multicolumn{3}{l}{} & \multicolumn{4}{l}{Mass Fractions (top 20 \textit{bd} nuclei)} & \multicolumn{1}{l}{}         \\
				\hline
				{}      & \multicolumn{2}{c}{$T_9$ = 7.33}     & \multicolumn{2}{c}{$T_9$ = 5.39}     & \multicolumn{2}{c}{$T_9$ = 4.93}     & \multicolumn{2}{c}{$T_9$ = 4.24}     \\
				{} & \multicolumn{2}{c}{$\rho$ = 4.01(+10)} & \multicolumn{2}{c}{$\rho$ = 2.20(+09)} & \multicolumn{2}{c}{$\rho$ = 1.06(+09)} & \multicolumn{2}{c}{$\rho$ = 3.30(+08)} \\
				{}      & \multicolumn{2}{c}{$Y_e$ = 0.41}     & \multicolumn{2}{c}{$Y_e$ = 0.425}    & \multicolumn{2}{c}{$Y_e$ = 0.43}     & \multicolumn{2}{c}{$Y_e$ = 0.44}     \\ 
				& pn-QRPA       & IPM        & pn-QRPA       & IPM        & pn-QRPA       & IPM       & pn-QRPA       & IPM        \\
				\hline
				\hline
				$^{49}$Sc & 5.44(-03) & 1.39(-02) & 9.28(-03) & ---       & 6.65(-03) & ---       & 1.18(-03) & ---      \\
				$^{65}$Co & 3.96(-03) & ---       & 1.04(-03) & 3.50(-03) & 2.90(-04) & 9.56(-04) & 2.93(-06) & 9.43(-06) \\
				$^{63}$Co & 2.35(-03) & 4.78(-03) & 6.26(-03) & 1.89(-02) & 5.15(-03) & 1.69(-02) & 9.42(-04) & 3.34(-03) \\
				$^{50}$Sc & 8.34(-03) & 1.03(-02) & 2.58(-03) & 4.66(-03) & 8.86(-04) & 1.74(-03) & 2.59(-05) & 5.81(-05) \\
				$^{59}$Mn & 2.09(-03) & 3.16(-03) & 1.90(-03) & 3.38(-03) & 9.82(-04) & 1.76(-03) & 5.97(-05) & 1.08(-04) \\
				$^{53}$Mn & 6.45(-10) & ---       & 4.12(-09) & ---       & 9.54(-09) & ---       & 1.21(-07) & 2.66(-07) \\
				$^{64}$Co & 3.29(-03) & 2.56(-03) & 1.80(-03) & 1.44(-03) & 7.14(-04) & 5.68(-04) & 2.13(-05) & 1.69(-05) \\
				$^{49}$Ca & 3.77(-02) & 3.88(-02) & 5.84(-03) & 6.00(-03) & 1.29(-03) & ---       & 9.47(-06) & 8.52(-06) \\
				$^{55}$Mn & 6.03(-06) & ---       & 9.36(-05) & ---       & 2.25(-04) & 4.85(-04) & 1.34(-03) & 3.62(-03) \\
				$^{58}$Cr & 3.44(-03) & ---       & 1.08(-03) & 1.78(-03) & 3.15(-04) & 4.99(-04) & 3.86(-06) & 5.98(-06) \\
				$^{61}$Fe & 3.43(-03) & 6.09(-03) & 6.37(-03) & 1.47(-02) & 4.30(-03) & 1.06(-02) & 4.83(-04) & 1.29(-03) \\
				$^{51}$Ti & 8.77(-03) & 8.77(-03) & 2.64(-02) & ---       & 2.43(-02) & 2.91(-02) & 7.60(-03) & 9.99(-03) \\
				$^{51}$Sc & 2.93(-03) & 1.08(-02) & 3.53(-04) & 1.51(-03) & 7.49(-05) & ---       & 5.68(-07) & 2.37(-06) \\
				$^{57}$Fe & 1.76(-06) & ---       & 2.89(-05) & 2.66(-05) & 7.44(-05) & ---       & 5.65(-04) & 8.47(-04) \\
				$^{57}$Cr & 2.63(-03) & 5.92(-03) & 1.51(-03) & 4.37(-03) & 6.38(-04) & 1.91(-03) & 2.40(-05) & 7.66(-05) \\
				$^{55}$Cr & 1.37(-03) & ---       & 6.96(-03) & ---       & 8.35(-03) & 1.78(-02) & 5.29(-03) & 1.29(-02) \\
				$^{62}$Fe & 1.19(-02) & ---       & 1.68(-02) & 1.62(-02) & 9.15(-03) & 8.38(-03) & 4.42(-04) & 3.80(-04) \\
				$^{57}$Mn & 4.50(-04) & 6.87(-04) & 2.84(-03) & 5.92(-03) & 3.78(-03) & 8.95(-03) & 3.21(-03) & 8.60(-03) \\
				$^{57}$Co & 3.27(-11) & ---       & 2.25(-10) & ---       & 5.79(-10) & ---       & 1.10(-08) & ---      \\
				$^{53}$Ti & 6.09(-03) & 1.28(-02) & 1.63(-03) & 3.86(-03) & 4.88(-04) & ---       & 7.97(-06) & 1.73(-05) \\
				\hline
		\end{tabular}}
	\end{table*}
	
	\begin{table*}[]
		\caption{ Same as Table \ref{T1MFbd} but with different $T_9$, $\rho$ and $Y_e$ values.}
		\label{T2MFbd}
		\centering
		\scalebox{0.7}{
			\begin{tabular}{c|cc|cc|cc|cc}
				
				\hline
				\multicolumn{3}{l}{} & \multicolumn{4}{l}{Mass Fractions (top 20 \textit{bd} nuclei)} & \multicolumn{1}{l}{}         \\
				\hline
				{}     & \multicolumn{2}{c}{$T_9$ = 3.8}      & \multicolumn{2}{c}{$T_9$ = 3.65}     & \multicolumn{2}{c}{$T_9$ = 3.4}      & \multicolumn{2}{c}{$T_9$ = 3.26}     \\
				{} & \multicolumn{2}{c}{$\rho$ = 1.45(+08)} & \multicolumn{2}{c}{$\rho$ = 1.07(+08)} & \multicolumn{2}{c}{$\rho$ = 5.86(+07)} & \multicolumn{2}{c}{$\rho$ = 4.32(+07)} \\
				{}      & \multicolumn{2}{c}{$Y_e$ = 0.45}     & \multicolumn{2}{c}{$Y_e$ = 0.455}    & \multicolumn{2}{c}{$Y_e$ = 0.47}     & \multicolumn{2}{c}{$Y_e$ = 0.485}    \\
				& pn-QRPA & IPM        & pn-QRPA       & IPM       & pn-QRPA       & IPM        & pn-QRPA       & IPM        \\
				\hline
				\hline
				$^{49}$Sc & 5.71(-06) & ---       & 4.18(-07) & ---       & 1.36(-14) & ---      & 1.62(-23) & ---      \\
				$^{65}$Co & 1.93(-11) & ---       & 1.01(-13) & ---       & 1.97(-27) & ---      & 1.99(-42) & ---      \\
				$^{63}$Co & 1.60(-06) & 6.93(-06) & 6.71(-08) & 2.66(-07) & 2.23(-17) & ---      & 9.48(-29) & ---      \\
				$^{50}$Sc & 5.91(-09) & ---       & 1.37(-10) & ---       & 2.81(-20) & ---      & 4.55(-31) & ---      \\
				$^{59}$Mn & 2.04(-08) & ---       & 4.93(-10) & ---       & 2.72(-20) & ---      & 4.45(-32) & ---      \\
				$^{53}$Mn & 4.34(-05) & 8.08(-05) & 1.82(-04) & 3.74(-04) & 6.39(-03) & 1.30(-02)& 2.18(-04) & 4.86(-04) \\
				$^{64}$Co & 1.67(-09) & ---       & 2.20(-11) & ---       & 4.52(-23) & ---      & 2.59(-36) & ---      \\
				$^{49}$Ca & 9.83(-11) & ---       & 7.37(-13) & ---       & 6.61(-25) & ---      & 1.27(-37) & ---      \\
				$^{55}$Mn & 1.12(-02) & 2.94(-02) & 1.18(-02) & 3.21(-02) & 9.08(-05) & 9.08(-05)& 1.68(-09) & 5.51(-09) \\
				$^{58}$Cr & 5.05(-11) & ---       & 3.70(-13) & ---       & 7.98(-26) & ---      & 1.44(-39) & ---      \\
				$^{61}$Fe & 3.86(-07) & 1.31(-06) & 1.26(-08) & ---       & 1.78(-18) & ---      & 4.82(-30) & ---      \\
				$^{51}$Ti & 8.32(-05) & 1.28(-04) & 8.08(-06) & 1.18(-05) & 6.49(-13) & ---      & 1.26(-21) & ---      \\
				$^{51}$Sc & 8.65(-12) & ---       & 7.25(-14) & ---       & 1.18(-25) & ---      & 3.03(-38) & ---      \\
				$^{57}$Fe & 8.10(-03) & 1.19(-02) & 1.02(-02) & 1.56(-02) & 1.61(-04) & 1.61(-04)& 4.26(-09) & 7.92(-09) \\
				$^{57}$Cr & 3.93(-09) & ---       & 7.41(-11) & ---       & 1.75(-21) & ---      & 1.84(-33) & ---      \\
				$^{55}$Cr & 2.08(-04) & 5.85(-04) & 3.08(-05) & 8.36(-05) & 1.18(-11) & ---      & 5.10(-20) & ---      \\
				$^{62}$Fe & 3.62(-08) & ---       & 5.06(-10) & ---       & 7.80(-22) & ---      & 4.05(-35) & ---      \\
				$^{57}$Mn & 2.27(-04) & 6.72(-04) & 4.06(-05) & 1.16(-04) & 3.26(-11) & 1.03(-10)& 2.04(-19) & ---      \\
				$^{57}$Co & 1.10(-05) & 3.84(-05) & 6.44(-05) & 2.53(-04) & 9.01(-03) & 3.55(-02)& 6.12(-04) & 2.69(-03) \\
				$^{53}$Ti & 3.25(-10) & ---       & 3.85(-12) & ---       & 1.77(-23) & ---      & 7.86(-36) & ---    \\
				\hline 
		\end{tabular}}
	\end{table*}

	\begin{table*}[]
			\tiny
		\caption{ The comparison of computed \textit{ec} rates ($\lambda^{ec}$) for $^{56}$Mn, $^{52}$V, $^{60}$Co, $^{53}$Mn, $^{49}$Sc, $^{50}$Sc, $^{55}$Fe, $^{59}$Co, $^{61}$Ni and $^{51}$Ti, with LSSM calculation \cite{Lan01}, as a function of $\rho$$\it Y_{e}$ and $T_9$ values. The units of $T_9$, $\rho$$\it Y_{e}$ and $\lambda^{ec}$ are \textit{GK}, $g/cm^3$ and $s^{-1}$ respectively.}
		\label{T1ec}
		\centering
		\scalebox{1.4}{
			\begin{tabular}{c|c|cc|cc|cc|cc}
				
				\hline
				\multicolumn{4}{l}{} & \multicolumn{2}{l}{$\lambda^{ec}$} & \multicolumn{4}{l}{}         \\
				\cline{1-10} \multicolumn{1}{l}{} &  & \multicolumn{2}{c}{$\rho$$\it Y_{e}$ = $10^7$ }        & \multicolumn{2}{c}{$\rho$$\it Y_{e}$ = $10^8$ } & \multicolumn{2}{c}{$\rho$$\it Y_{e}$ = $10^9$ }       & \multicolumn{2}{c}{$\rho$$\it Y_{e}$ = $10^{10}$ }       \\
				\cline{3-10} \multicolumn{1}{l}{Nuclei} &       $T_9$                     & pn-QRPA   & LSSM      & pn-QRPA   & LSSM      & pn-QRPA  & LSSM      & pn-QRPA  & LSSM     \\
				\hline
				\hline
				{$^{56}$Mn} & 1  & 1.17(-26) & 2.52(-11) & 1.82(-20) & 8.41(-06) & 3.27(-07) & 1.37(-02) & 5.36(+01) & 2.79(+01) \\
				& 3  & 1.67(-10) & 4.16(-07) & 2.77(-08) & 4.66(-05) & 5.22(-04) & 1.91(-02) & 9.18(+01) & 3.10(+01) \\
				& 10 & 4.53(-03) & 1.49(-03) & 2.03(-02) & 6.44(-03) & 7.53(-01) & 2.02(-01) & 2.98(+02) & 4.78(+01) \\
				& 30 & 2.34(+02) & 1.04(+01) & 2.51(+02) & 1.12(+01) & 4.94(+02) & 2.19(+01) & 7.13(+03) & 2.90(+02) \\
				\hline
				{$^{52}$V}  & 1  & 1.71(-26) & 4.52(-13) & 2.92(-20) & 6.04(-07) & 1.60(-06) & 8.81(-03) & 3.76(+03) & 1.28(+01) \\
				& 3  & 3.61(-09) & 1.54(-07) & 6.27(-07) & 2.24(-05) & 2.77(-02) & 1.59(-02) & 3.73(+03) & 1.42(+01) \\
				& 10 & 2.56(-01) & 1.07(-03) & 1.16(+00) & 4.58(-03) & 4.74(+01) & 1.29(-01) & 1.58(+04) & 2.44(+01) \\
				& 30 & 3.42(+03) & 7.98(+00) & 3.67(+03) & 8.57(+00) & 7.23(+03) & 1.66(+01) & 1.04(+05) & 2.13(+02) \\
				\hline
				{$^{60}$Co} & 1  & 3.70(-19) & 2.01(-09) & 4.91(-13) & 8.73(-07) & 4.71(-03) & 7.19(-03) & 7.48(+01) & 5.92(+01) \\
				& 3  & 2.11(-08) & 1.17(-06) & 2.91(-06) & 4.35(-05) & 1.94(-02) & 1.83(-02) & 1.14(+02) & 6.43(+01) \\
				& 10 & 8.77(-03) & 3.40(-03) & 3.87(-02) & 1.45(-02) & 1.27(+00) & 4.32(-01) & 3.40(+02) & 9.25(+01) \\
				& 30 & 3.51(+02) & 1.72(+01) & 3.77(+02) & 1.85(+01) & 7.41(+02) & 3.59(+01) & 1.07(+04) & 4.70(+02) \\
				\hline
				{$^{53}$Mn} & 1  & 3.21(-05) & 2.10(-06) & 1.01(-03) & 5.85(-04) & 4.74(-01) & 1.67(-01) & 1.35(+02) & 1.52(+02) \\
				& 3  & 7.11(-05) & 2.15(-05) & 2.22(-03) & 1.01(-03) & 7.40(-01) & 2.39(-01) & 1.69(+02) & 1.58(+02) \\
				& 10 & 4.31(-02) & 1.22(-02) & 1.76(-01) & 5.25(-02) & 3.76(+00) & 1.58(+00) & 2.75(+02) & 2.06(+02) \\
				& 30 & 4.76(+02) & 2.56(+01) & 5.11(+02) & 2.75(+01) & 9.84(+02) & 5.37(+01) & 1.16(+04) & 7.06(+02) \\
				\hline
				{$^{49}$Sc} & 1  & 1.52(-26) & 2.34(-29) & 2.59(-20) & 2.46(-23) & 6.82(-07) & 4.72(-18) & 1.08(+02) & 3.82(-01) \\
				& 3  & 2.92(-09) & 1.67(-12) & 5.07(-07) & 1.67(-10) & 2.00(-02) & 1.42(-07) & 2.05(+02) & 5.08(-01) \\
				& 10 & 3.63(-02) & 1.36(-04) & 1.63(-01) & 5.65(-04) & 6.43(+00) & 1.25(-02) & 1.18(+03) & 2.74(+00) \\
				& 30 & 4.06(+02) & 4.25(+00) & 4.37(+02) & 4.55(+00) & 8.55(+02) & 8.75(+00) & 1.16(+04) & 1.02(+02) \\
				\hline
				{$^{50}$Sc} & 1  & 1.04(-30) & 1.24(-27) & 1.76(-24) & 1.98(-21) & 1.12(-10) & 3.38(-12) & 1.22(+02) & 8.85(-02) \\
				& 3  & 3.83(-11) & 5.19(-12) & 6.65(-09) & 7.10(-10) & 3.15(-04) & 5.11(-06) & 1.81(+02) & 2.02(-01) \\
				& 10 & 8.83(-03) & 7.94(-05) & 3.97(-02) & 3.39(-04) & 1.64(+00) & 8.81(-03) & 7.59(+02) & 1.93(+00) \\
				& 30 & 6.68(+02) & 2.79(+00) & 7.18(+02) & 3.00(+00) & 1.42(+03) & 5.78(+00) & 2.11(+04) & 7.06(+01) \\
				\hline
				{$^{55}$Fe} & 1  & 8.41(-05) & 1.20(-05) & 4.13(-03) & 5.61(-04) & 2.47(-01) & 3.21(-01) & 8.30(+01) & 2.14(+02) \\
				& 3  & 2.33(-04) & 4.72(-05) & 6.30(-03) & 1.54(-03) & 3.16(-01) & 4.47(-01) & 8.85(+01) & 2.21(+02) \\
				& 10 & 2.04(-02) & 2.14(-02) & 8.38(-02) & 9.12(-02) & 1.75(+00) & 2.59(+00) & 1.69(+02) & 2.87(+02) \\
				& 30 & 1.91(+02) & 3.47(+01) & 2.05(+02) & 3.72(+01) & 4.00(+02) & 7.24(+01) & 5.37(+03) & 9.33(+02) \\
				\hline
				{$^{59}$Co} & 1  & 3.27(-11) & 6.82(-13) & 7.21(-06) & 9.33(-08) & 6.14(-02) & 8.97(-03) & 1.19(+02) & 8.53(+01) \\
				& 3  & 1.34(-06) & 1.72(-07) & 1.41(-04) & 1.55(-05) & 1.04(-01) & 1.64(-02) & 1.19(+02) & 8.81(+01) \\
				& 10 & 8.02(-03) & 4.11(-03) & 3.48(-02) & 1.77(-02) & 1.01(+00) & 5.53(-01) & 2.00(+02) & 1.22(+02) \\
				& 30 & 2.82(+02) & 2.10(+01) & 3.03(+02) & 2.25(+01) & 5.94(+02) & 4.39(+01) & 8.26(+03) & 5.85(+02) \\
				\hline
				{$^{61}$Ni} & 1  & 7.69(-10) & 1.52(-09) & 8.36(-05) & 8.41(-05) & 8.00(-02) & 6.68(-02) & 7.16(+01) & 1.38(+02) \\
				& 3  & 6.14(-06) & 5.04(-06) & 5.89(-04) & 3.92(-04) & 1.26(-01) & 1.07(-01) & 8.43(+01) & 1.46(+02) \\
				& 10 & 7.35(-03) & 6.84(-03) & 3.15(-02) & 2.99(-02) & 8.26(-01) & 9.93(-01) & 1.73(+02) & 1.79(+02) \\
				& 30 & 2.30(+02) & 2.40(+01) & 2.48(+02) & 2.58(+01) & 4.86(+02) & 5.02(+01) & 6.90(+03) & 6.71(+02) \\
				\hline
				{$^{51}$Ti} & 1  & 9.64(-34) & 3.93(-34) & 1.64(-27) & 5.28(-28) & 1.01(-13) & 7.53(-15) & 1.37(+00) & 2.42(-01) \\
				& 3  & 1.77(-12) & 2.94(-13) & 3.08(-10) & 4.09(-11) & 1.45(-05) & 6.64(-07) & 1.67(+01) & 3.50(-01) \\
				& 10 & 1.44(-03) & 1.25(-04) & 6.49(-03) & 5.28(-04) & 2.67(-01) & 1.34(-02) & 1.11(+02) & 2.96(+00) \\
				& 30 & 1.89(+02) & 3.17(+00) & 2.03(+02) & 3.40(+00) & 4.00(+02) & 6.58(+00) & 5.75(+03) & 8.17(+01) \\
				\hline
		\end{tabular}}
	\end{table*}
	
	\begin{table*}[]
			\tiny
		\caption{ Same as Table \ref{T1ec} but for $^{56}$Ni, $^{56}$Fe, $^{65}$Cu, $^{55}$Co, $^{57}$Ni, $^{54}$Cr, $^{64}$Cu, $^{57}$Co, $^{63}$Cu and $^{58}$Ni.}
		\label{T2ec}
		\centering
		\scalebox{1.4}{
			\begin{tabular}{c|c|cc|cc|cc|cc}
				
				\hline
				\multicolumn{4}{l}{} & \multicolumn{2}{l}{$\lambda^{ec}$ } & \multicolumn{4}{l}{}         \\
				\cline{1-10} \multicolumn{1}{l}{} &  & \multicolumn{2}{c}{$\rho$$\it Y_{e}$ = $10^7$ }        & \multicolumn{2}{c}{$\rho$$\it Y_{e}$ = $10^8$ } & \multicolumn{2}{c}{$\rho$$\it Y_{e}$ = $10^9$ }       & \multicolumn{2}{c}{$\rho$$\it Y_{e}$ = $10^{10}$ }       \\
				\cline{3-10} \multicolumn{1}{l}{Nuclei} &       $T_9$                     & pn-QRPA   & LSSM      & pn-QRPA   & LSSM      & pn-QRPA  & LSSM      & pn-QRPA  & LSSM     \\
				\hline
				\hline
				{$^{56}$Ni} & 1  & 3.39(-04) & 8.11(-04) & 1.33(-02) & 5.31(-02) & 1.36(+00) & 1.18(+01) & 7.74(+02) & 1.08(+03) \\
				& 3  & 5.94(-04) & 1.87(-03) & 1.58(-02) & 9.14(-02) & 1.80(+00) & 1.27(+01) & 7.89(+02) & 1.10(+03) \\
				& 10 & 9.89(-02) & 2.54(-01) & 4.13(-01) & 1.07(+00) & 1.10(+01) & 2.43(+01) & 1.03(+03) & 1.21(+03) \\
				& 30 & 6.47(+02) & 1.04(+02) & 6.95(+02) & 1.11(+02) & 1.35(+03) & 2.15(+02) & 1.66(+04) & 2.63(+03) \\
				\hline
				{$^{56}$Fe} & 1  & 1.66(-21) & 1.92(-20) & 2.82(-15) & 2.39(-14) & 6.05(-03) & 8.77(-03) & 6.58(+01) & 8.67(+01) \\
				& 3  & 3.19(-09) & 1.24(-08) & 5.28(-07) & 1.77(-06) & 1.57(-02) & 2.40(-02) & 6.90(+01) & 9.25(+01) \\
				& 10 & 3.60(-03) & 4.85(-03) & 1.56(-02) & 2.11(-02) & 4.93(-01) & 6.95(-01) & 1.19(+02) & 1.34(+02) \\
				& 30 & 9.57(+01) & 2.22(+01) & 1.03(+02) & 2.38(+01) & 2.01(+02) & 4.63(+01) & 2.76(+03) & 6.00(+02) \\
				\hline
				{$^{65}$Cu} & 1  & 1.10(-12) & 2.91(-13) & 4.25(-07) & 1.96(-07) & 3.30(-01) & 5.81(-02) & 1.44(+02) & 3.43(+01) \\
				& 3  & 3.81(-06) & 9.16(-07) & 4.30(-04) & 8.63(-05) & 4.16(-01) & 6.82(-02) & 1.67(+02) & 3.57(+01) \\
				& 10 & 3.33(-02) & 5.25(-03) & 1.44(-01) & 2.19(-02) & 3.86(+00) & 5.08(-01) & 4.26(+02) & 6.00(+01) \\
				& 30 & 7.31(+02) & 1.56(+01) & 7.85(+02) & 1.68(+01) & 1.53(+03) & 3.24(+01) & 2.03(+04) & 4.01(+02) \\
				\hline
				{$^{55}$Co} & 1  & 5.74(-03) & 4.62(-04) & 1.09(-01) & 1.60(-02) & 4.42(+00) & 4.57(+00) & 6.93(+02) & 6.73(+02) \\
				& 3  & 8.17(-03) & 7.48(-04) & 1.40(-01) & 2.49(-02) & 5.41(+00) & 5.07(+00) & 8.00(+02) & 6.81(+02) \\
				& 10 & 1.91(-01) & 1.11(-01) & 7.80(-01) & 4.74(-01) & 1.65(+01) & 1.24(+01) & 1.10(+03) & 7.91(+02) \\
				& 30 & 7.87(+02) & 7.14(+01) & 8.45(+02) & 7.67(+01) & 1.63(+03) & 1.49(+02) & 1.97(+04) & 1.86(+03) \\
				\hline
				{$^{57}$Ni} & 1  & 7.62(-03) & 6.92(-04) & 1.11(-01) & 2.44(-02) & 2.98(+00) & 7.26(+00) & 2.21(+02) & 8.97(+02) \\
				& 3  & 1.50(-02) & 1.32(-03) & 2.07(-01) & 4.35(-02) & 4.92(+00) & 7.82(+00) & 3.04(+02) & 8.97(+02) \\
				& 10 & 1.57(-01) & 1.50(-01) & 6.21(-01) & 6.41(-01) & 1.03(+01) & 1.65(+01) & 5.46(+02) & 9.89(+02) \\
				& 30 & 4.74(+02) & 9.16(+01) & 5.09(+02) & 9.84(+01) & 9.91(+02) & 1.91(+02) & 1.30(+04) & 2.37(+03) \\
				\hline
				{$^{54}$Cr} & 1  & 1.92(-35) & 2.11(-36) & 3.24(-29) & 3.00(-30) & 2.05(-15) & 1.82(-17) & 5.30(+01) & 5.25(+00) \\
				& 3  & 7.93(-13) & 1.93(-13) & 1.37(-10) & 2.86(-11) & 6.35(-06) & 5.66(-07) & 5.62(+01) & 5.94(+00) \\
				& 10 & 1.11(-03) & 2.59(-04) & 4.98(-03) & 1.12(-03) & 2.00(-01) & 3.48(-02) & 1.05(+02) & 1.41(+01) \\
				& 30 & 7.33(+01) & 7.00(+00) & 7.89(+01) & 7.52(+00) & 1.55(+02) & 1.46(+01) & 2.25(+03) & 1.85(+02) \\
				\hline
				{$^{64}$Cu} & 1  & 1.11(-10) & 9.23(-04) & 6.12(-06) & 1.84(-02) & 3.83(-01) & 6.90(-01) & 3.17(+02) & 1.35(+02) \\
				& 3  & 1.51(-05) & 5.96(-04) & 1.29(-03) & 1.17(-02) & 9.12(-01) & 5.24(-01) & 5.11(+02) & 1.29(+02) \\
				& 10 & 9.48(-02) & 1.33(-02) & 4.06(-01) & 5.61(-02) & 1.08(+01) & 1.40(+00) & 1.19(+03) & 1.62(+02) \\
				& 30 & 8.11(+02) & 1.98(+01) & 8.71(+02) & 2.13(+01) & 1.70(+03) & 4.16(+01) & 2.28(+04) & 5.61(+02) \\
				\hline
				{$^{57}$Co} & 1  & 1.86(-05) & 1.27(-05) & 2.03(-03) & 5.18(-04) & 6.07(-01) & 2.80(-01) & 2.79(+02) & 2.70(+02) \\
				& 3  & 1.38(-04) & 2.90(-05) & 5.37(-03) & 9.51(-04) & 7.00(-01) & 3.85(-01) & 2.70(+02) & 2.76(+02) \\
				& 10 & 3.18(-02) & 2.42(-02) & 1.33(-01) & 1.02(-01) & 3.21(+00) & 2.94(+00) & 3.78(+02) & 3.51(+02) \\
				& 30 & 3.87(+02) & 4.36(+01) & 4.17(+02) & 4.68(+01) & 8.11(+02) & 9.10(+01) & 1.06(+04) & 1.17(+03) \\
				\hline
				{$^{63}$Cu} & 1  & 3.11(-06) & 5.74(-06) & 2.18(-03) & 1.28(-03) & 1.13(+00) & 2.28(-01) & 3.48(+02) & 1.08(+02) \\
				& 3  & 1.96(-04) & 4.20(-05) & 1.05(-02) & 2.18(-03) & 1.64(+00) & 2.48(-01) & 4.07(+02) & 1.12(+02) \\
				& 10 & 9.31(-02) & 9.68(-03) & 3.93(-01) & 4.10(-02) & 9.59(+00) & 1.09(+00) & 7.71(+02) & 1.60(+02) \\
				& 30 & 1.02(+03) & 2.22(+01) & 1.10(+03) & 2.39(+01) & 2.13(+03) & 4.66(+01) & 2.69(+04) & 6.30(+02) \\
				\hline
				{$^{58}$Ni} & 1  & 6.12(-06) & 2.85(-10) & 3.81(-03) & 7.18(-05) & 5.82(-01) & 9.10(-01) & 3.44(+02) & 4.06(+02) \\
				& 3  & 8.95(-05) & 7.35(-06) & 5.65(-03) & 6.95(-04) & 6.37(-01) & 1.14(+00) & 3.52(+02) & 4.13(+02) \\
				& 10 & 3.01(-02) & 4.30(-02) & 1.28(-01) & 1.85(-01) & 3.46(+00) & 5.46(+00) & 4.76(+02) & 5.19(+02) \\
				& 30 & 2.86(+02) & 5.74(+01) & 3.08(+02) & 6.17(+01) & 6.00(+02) & 1.20(+02) & 7.83(+03) & 1.51(+03)\\
				\hline
		\end{tabular}}
	\end{table*}
	
	
	\begin{table*}[]
			\tiny
		\caption{ The comparison of computed \textit{bd} rates ($\lambda^{bd}$) for $^{49}$Sc, $^{65}$Co, $^{63}$Co, $^{50}$Sc, $^{59}$Mn, $^{53}$Mn, $^{64}$Co, $^{49}$Ca, $^{55}$Mn and $^{58}$Cr, with LSSM calculation \cite{Lan01}, as a function of $\rho$$\it Y_{e}$ and $T_9$ values. The units of $T_9$, $\rho$$\it Y_{e}$ and $\lambda^{bd}$ are \textit{GK}, $g/cm^3$ and $s^{-1}$ respectively.}
		\label{T1bd}
		\centering
		\scalebox{1.4}{
			\begin{tabular}{c|c|cc|cc|cc|cc}
				
				\hline
				\multicolumn{4}{l}{} & \multicolumn{2}{l}{$\lambda^{bd}$ } & \multicolumn{4}{l}{}         \\
				\cline{1-10} \multicolumn{1}{l}{} &  & \multicolumn{2}{c}{$\rho$$\it Y_{e}$ = $10^7$ }        & \multicolumn{2}{c}{$\rho$$\it Y_{e}$ = $10^8$ } & \multicolumn{2}{c}{$\rho$$\it Y_{e}$ = $10^9$ }       & \multicolumn{2}{c}{$\rho$$\it Y_{e}$ = $10^{10}$ }       \\
				\cline{3-10} \multicolumn{1}{l}{Nuclei} &       $T_9$                     & pn-QRPA   & LSSM      & pn-QRPA   & LSSM      & pn-QRPA  & LSSM      & pn-QRPA  & LSSM     \\
				\hline
				\hline
				{$^{49}$Sc} & 1  & 4.76(-03) & 1.23(-04) & 4.36(-05) & 8.91(-07) & 1.28(-17) & 2.18(-20) & 1.55(-47) & 2.32(-47) \\
				& 3  & 8.07(-02) & 1.42(-04) & 1.58(-02) & 1.12(-05) & 1.26(-06) & 3.59(-09) & 1.26(-16) & 4.74(-17) \\
				& 10 & 7.10(-01) & 1.05(-02) & 4.72(-01) & 9.46(-03) & 3.16(-02) & 3.80(-03) & 2.69(-05) & 1.77(-05) \\
				& 30 & 3.06(+00) & 1.69(-01) & 3.01(+00) & 1.66(-01) & 2.36(+00) & 1.43(-01) & 3.59(-01) & 3.14(-02) \\
				\hline
				{$^{65}$Co} & 1  & 6.75(-01) & 5.14(-01) & 5.00(-01) & 3.79(-01) & 3.89(-02) & 2.89(-02) & 7.83(-28) & 3.50(-27) \\
				& 3  & 9.12(-01) & 5.45(-01) & 6.90(-01) & 4.10(-01) & 7.45(-02) & 4.37(-02) & 8.38(-11) & 4.52(-10) \\
				& 10 & 1.70(+00) & 1.72(+00) & 1.48(+00) & 1.56(+00) & 4.20(-01) & 6.55(-01) & 8.18(-04) & 3.01(-03) \\
				& 30 & 1.54(+01) & 2.08(+00) & 1.52(+01) & 2.06(+00) & 1.25(+01) & 1.76(+00) & 2.25(+00) & 3.89(-01) \\
				\hline
				{$^{63}$Co} & 1  & 3.38(-02) & 2.24(-02) & 1.23(-02) & 8.39(-03) & 6.12(-10) & 1.91(-10) & 8.09(-40) & 8.79(-39) \\
				& 3  & 6.52(-02) & 2.49(-02) & 3.05(-02) & 1.10(-02) & 3.81(-05) & 5.24(-05) & 4.55(-15) & 8.00(-14) \\
				& 10 & 1.60(-01) & 3.03(-01) & 1.23(-01) & 2.65(-01) & 1.56(-02) & 8.49(-02) & 1.74(-05) & 2.24(-04) \\
				& 30 & 2.22(+00) & 9.71(-01) & 2.18(+00) & 9.57(-01) & 1.74(+00) & 8.07(-01) & 2.69(-01) & 1.64(-01) \\
				\hline
				{$^{50}$Sc} & 1  & 6.81(-02) & 7.05(-03) & 5.12(-03) & 3.09(-03) & 6.12(-15) & 2.33(-05) & 5.09(-44) & 8.99(-24) \\
				& 3  & 1.15(+00) & 1.22(-02) & 2.28(-01) & 7.06(-03) & 1.69(-05) & 4.95(-04) & 1.79(-15) & 5.73(-10) \\
				& 10 & 4.45(+00) & 1.72(-01) & 2.98(+00) & 1.58(-01) & 2.19(-01) & 7.57(-02) & 1.93(-04) & 5.90(-04) \\
				& 30 & 1.96(+01) & 6.68(-01) & 1.91(+01) & 6.61(-01) & 1.44(+01) & 5.74(-01) & 1.76(+00) & 1.39(-01) \\
				\hline
				{$^{59}$Mn} & 1  & 1.95(-01) & 9.79(-02) & 7.91(-02) & 6.05(-02) & 4.33(-04) & 3.72(-04) & 2.84(-32) & 3.26(-31) \\
				& 3  & 2.59(-01) & 1.01(-01) & 1.26(-01) & 6.49(-02) & 2.94(-03) & 1.59(-03) & 1.81(-12) & 1.59(-11) \\
				& 10 & 1.06(+00) & 4.99(-01) & 8.61(-01) & 4.50(-01) & 1.69(-01) & 1.86(-01) & 3.33(-04) & 1.14(-03) \\
				& 30 & 1.07(+01) & 1.99(+00) & 1.05(+01) & 1.97(+00) & 8.85(+00) & 1.71(+00) & 1.93(+00) & 4.25(-01) \\
				\hline
				{$^{53}$Mn} & 1  & 6.67(-31) & 7.80(-29) & 3.67(-34) & 1.11(-33) & 8.18(-48) & 1.75(-46) & 9.42(-78) & 3.58(-76) \\
				& 3  & 2.50(-12) & 6.47(-12) & 1.29(-13) & 4.99(-13) & 3.44(-18) & 7.28(-17) & 3.33(-28) & 1.21(-26) \\
				& 10 & 3.05(-05) & 1.09(-04) & 2.03(-05) & 7.93(-05) & 2.73(-06) & 9.82(-06) & 1.14(-08) & 1.40(-08) \\
				& 30 & 4.93(-01) & 1.96(-02) & 4.86(-01) & 1.92(-02) & 4.17(-01) & 1.57(-02) & 9.42(-02) & 2.99(-03) \\
				\hline
				{$^{64}$Co} & 1  & 1.63(-01) & 1.73(+00) & 7.96(-02) & 1.42(+00) & 1.50(-06) & 3.40(-01) & 2.54(-36) & 4.47(-21) \\
				& 3  & 4.05(-01) & 8.07(-01) & 2.30(-01) & 6.56(-01) & 1.47(-03) & 1.55(-01) & 2.01(-13) & 5.87(-09) \\
				& 10 & 9.62(-01) & 8.61(-01) & 7.71(-01) & 7.80(-01) & 1.24(-01) & 3.24(-01) & 1.52(-04) & 1.60(-03) \\
				& 30 & 5.35(+00) & 1.35(+00) & 5.24(+00) & 1.33(+00) & 4.17(+00) & 1.14(+00) & 6.19(-01) & 2.56(-01) \\
				\hline
				{$^{49}$Ca} & 1  & 3.05(-03) & 8.13(-04) & 1.75(-03) & 1.82(-05) & 1.09(-07) & 4.90(-17) & 3.44(-33) & 1.05(-30) \\
				& 3  & 1.26(-02) & 9.75(-04) & 7.60(-03) & 1.21(-04) & 8.05(-05) & 2.11(-06) & 8.95(-13) & 2.43(-11) \\
				& 10 & 3.48(-01) & 1.79(-01) & 3.13(-01) & 1.67(-01) & 1.22(-01) & 9.29(-02) & 3.81(-04) & 1.05(-03) \\
				& 30 & 1.27(+01) & 6.46(-01) & 1.25(+01) & 6.38(-01) & 1.07(+01) & 5.60(-01) & 2.48(+00) & 1.41(-01) \\
				\hline
				{$^{55}$Mn} & 1  & 8.13(-13) & 1.29(-13) & 1.28(-17) & 3.87(-17) & 2.96(-30) & 4.38(-29) & 7.11(-60) & 1.43(-58) \\
				& 3  & 6.18(-08) & 2.92(-07) & 5.74(-09) & 4.36(-08) & 2.82(-12) & 3.17(-11) & 8.09(-22) & 7.29(-21) \\
				& 10 & 8.75(-04) & 4.19(-03) & 7.01(-04) & 3.33(-03) & 1.41(-04) & 5.92(-04) & 2.08(-07) & 9.82(-07) \\
				& 30 & 4.12(-01) & 1.29(-01) & 4.05(-01) & 1.27(-01) & 3.25(-01) & 1.05(-01) & 5.09(-02) & 1.96(-02) \\
				\hline
				{$^{58}$Cr} & 1  & 1.11(-01) & 9.35(-02) & 4.35(-02) & 2.56(-02) & 6.71(-10) & 4.40(-09) & 5.47(-38) & 3.95(-36) \\
				& 3  & 1.14(-01) & 9.10(-02) & 4.94(-02) & 3.26(-02) & 3.14(-05) & 1.26(-04) & 1.06(-13) & 2.66(-12) \\
				& 10 & 5.15(-01) & 4.39(-01) & 4.48(-01) & 3.94(-01) & 1.59(-01) & 1.64(-01) & 6.78(-04) & 1.13(-03) \\
				& 30 & 8.30(+00) & 1.58(+00) & 8.18(+00) & 1.56(+00) & 7.00(+00) & 1.36(+00) & 1.57(+00) & 3.37(-01)\\
				\hline
		\end{tabular}}
	\end{table*}
	
	\begin{table*}[]
		\tiny
		\caption{ Same as Table \ref{T1bd} but for $^{61}$Fe, $^{51}$Ti, $^{51}$Sc, $^{57}$Fe, $^{57}$Cr, $^{55}$Cr, $^{62}$Fe, $^{57}$Mn, $^{57}$Co and $^{53}$Ti.}
		\label{T2bd}
		\centering
		\scalebox{1.4}{
			\begin{tabular}{c|c|cc|cc|cc|cc}
				
				\hline
				\multicolumn{4}{l}{} & \multicolumn{2}{l}{$\lambda^{bd}$ } & \multicolumn{4}{l}{}         \\
				\cline{1-10} \multicolumn{1}{l}{} &  & \multicolumn{2}{c}{$\rho$$\it Y_{e}$ = $10^7$ }        & \multicolumn{2}{c}{$\rho$$\it Y_{e}$ = $10^8$ } & \multicolumn{2}{c}{$\rho$$\it Y_{e}$ = $10^9$ }       & \multicolumn{2}{c}{$\rho$$\it Y_{e}$ = $10^{10}$ }       \\
				\cline{3-10} \multicolumn{1}{l}{Nuclei} &       $T_9$                     & pn-QRPA   & LSSM      & pn-QRPA   & LSSM      & pn-QRPA  & LSSM      & pn-QRPA  & LSSM     \\
				\hline
				\hline
				{$^{61}$Fe} & 1  & 6.14(-03) & 2.95(-03) & 1.80(-03) & 6.84(-04) & 8.95(-09) & 2.54(-09) & 1.01(-37) & 8.89(-37) \\
				& 3  & 2.87(-02) & 6.34(-03) & 1.34(-02) & 2.77(-03) & 7.91(-05) & 5.15(-05) & 3.76(-14) & 2.40(-13) \\
				& 10 & 1.15(-01) & 1.43(-01) & 9.40(-02) & 1.28(-01) & 2.24(-02) & 4.67(-02) & 5.96(-05) & 1.67(-04) \\
				& 30 & 7.46(+00) & 6.75(-01) & 7.36(+00) & 6.65(-01) & 6.21(+00) & 5.68(-01) & 1.26(+00) & 1.23(-01) \\
				\hline
				{$^{51}$Ti} & 1  & 1.80(-03) & 1.25(-03) & 7.14(-05) & 2.34(-05) & 2.47(-15) & 5.51(-16) & 1.32(-44) & 1.92(-44) \\
				& 3  & 3.36(-03) & 1.62(-03) & 7.08(-04) & 2.38(-04) & 1.64(-06) & 5.81(-07) & 3.55(-16) & 1.26(-15) \\
				& 10 & 1.07(-01) & 6.14(-02) & 8.91(-02) & 5.35(-02) & 1.89(-02) & 1.80(-02) & 2.40(-05) & 6.97(-05) \\
				& 30 & 4.05(+00) & 4.46(-01) & 3.98(+00) & 4.40(-01) & 3.21(+00) & 3.76(-01) & 5.02(-01) & 8.22(-02) \\
				\hline
				{$^{51}$Sc} & 1  & 6.71(-04) & 5.22(-02) & 3.96(-04) & 2.64(-02) & 4.79(-06) & 7.55(-05) & 1.72(-26) & 5.92(-25) \\
				& 3  & 5.11(-01) & 5.35(-02) & 3.50(-01) & 2.90(-02) & 1.74(-02) & 6.08(-04) & 8.18(-11) & 9.73(-10) \\
				& 10 & 5.38(+00) & 1.94(-01) & 4.55(+00) & 1.78(-01) & 1.06(+00) & 9.20(-02) & 1.70(-03) & 1.37(-03) \\
				& 30 & 1.07(+02) & 7.36(-01) & 1.05(+02) & 7.29(-01) & 8.63(+01) & 6.43(-01) & 1.51(+01) & 1.74(-01) \\
				\hline
				{$^{57}$Fe} & 1  & 1.86(-15) & 5.05(-16) & 2.34(-20) & 3.30(-20) & 5.69(-34) & 1.33(-31) & 6.53(-64) & 3.56(-61) \\
				& 3  & 2.38(-08) & 3.69(-08) & 8.81(-10) & 7.43(-09) & 2.68(-14) & 7.73(-12) & 2.61(-24) & 1.46(-21) \\
				& 10 & 9.62(-05) & 2.75(-03) & 7.06(-05) & 2.17(-03) & 9.20(-06) & 3.61(-04) & 1.05(-08) & 5.24(-07) \\
				& 30 & 1.48(-01) & 9.42(-02) & 1.45(-01) & 9.27(-02) & 1.16(-01) & 7.57(-02) & 1.76(-02) & 1.36(-02) \\
				\hline
				{$^{57}$Cr} & 1  & 1.19(-01) & 3.28(-02) & 7.55(-02) & 1.86(-02) & 2.87(-04) & 1.24(-04) & 8.05(-33) & 2.68(-31) \\
				& 3  & 1.40(-01) & 3.82(-02) & 9.25(-02) & 2.28(-02) & 2.48(-03) & 9.46(-04) & 1.58(-12) & 2.32(-11) \\
				& 10 & 3.94(-01) & 3.44(-01) & 3.37(-01) & 3.13(-01) & 9.12(-02) & 1.37(-01) & 1.77(-04) & 8.89(-04) \\
				& 30 & 6.84(+00) & 1.25(+00) & 6.75(+00) & 1.23(+00) & 5.61(+00) & 1.07(+00) & 1.06(+00) & 2.62(-01) \\
				\hline
				{$^{55}$Cr} & 1  & 1.11(-03) & 2.39(-03) & 1.04(-04) & 2.70(-04) & 5.35(-15) & 1.82(-15) & 9.57(-45) & 1.15(-43) \\
				& 3  & 4.84(-03) & 2.29(-03) & 1.43(-03) & 5.71(-04) & 1.09(-06) & 1.40(-06) & 1.57(-16) & 2.09(-15) \\
				& 10 & 4.84(-02) & 8.04(-02) & 3.82(-02) & 7.01(-02) & 6.27(-03) & 2.22(-02) & 7.26(-06) & 6.85(-05) \\
				& 30 & 1.45(+00) & 6.49(-01) & 1.42(+00) & 6.40(-01) & 1.15(+00) & 5.42(-01) & 1.88(-01) & 1.14(-01) \\
				\hline
				{$^{62}$Fe} & 1  & 1.10(-03) & 6.25(-03) & 6.27(-05) & 5.11(-05) & 9.71(-18) & 2.54(-15) & 9.25(-45) & 3.60(-43) \\
				& 3  & 1.25(-03) & 7.74(-03) & 1.68(-04) & 1.21(-03) & 1.49(-07) & 8.93(-06) & 6.00(-16) & 1.62(-14) \\
				& 10 & 1.34(-01) & 3.21(-01) & 1.12(-01) & 2.82(-01) & 3.24(-02) & 9.40(-02) & 1.58(-04) & 2.79(-04) \\
				& 30 & 5.47(+00) & 1.24(+00) & 5.40(+00) & 1.22(+00) & 4.59(+00) & 1.03(+00) & 1.04(+00) & 2.14(-01) \\
				\hline
				{$^{57}$Mn} & 1  & 5.48(-03) & 4.70(-03) & 7.78(-04) & 5.47(-04) & 4.01(-15) & 4.59(-15) & 1.49(-44) & 1.46(-43) \\
				& 3  & 9.23(-03) & 4.83(-03) & 2.32(-03) & 1.11(-03) & 3.15(-07) & 1.99(-06) & 1.11(-16) & 1.69(-15) \\
				& 10 & 2.98(-02) & 1.05(-01) & 2.14(-02) & 8.99(-02) & 2.88(-03) & 2.52(-02) & 7.05(-06) & 6.64(-05) \\
				& 30 & 2.23(+00) & 7.36(-01) & 2.20(+00) & 7.24(-01) & 1.83(+00) & 6.11(-01) & 3.47(-01) & 1.26(-01) \\
				\hline
				{$^{57}$Co} & 1  & 1.43(-28) & 1.71(-26) & 1.06(-33) & 5.20(-31) & 9.38(-47) & 2.74(-44) & 1.08(-76) & 4.63(-74) \\
				& 3  & 2.94(-13) & 7.41(-11) & 2.92(-14) & 4.81(-12) & 1.04(-18) & 3.06(-16) & 1.01(-28) & 4.38(-26) \\
				& 10 & 8.57(-07) & 2.01(-04) & 5.26(-07) & 1.36(-04) & 2.83(-08) & 1.30(-05) & 2.24(-11) & 1.77(-08) \\
				& 30 & 1.96(-03) & 1.82(-02) & 1.91(-03) & 1.79(-02) & 1.46(-03) & 1.46(-02) & 1.83(-04) & 2.79(-03) \\
				\hline
				{$^{53}$Ti} & 1  & 7.69(-03) & 1.34(-02) & 4.75(-03) & 3.89(-03) & 2.74(-05) & 4.00(-06) & 7.48(-32) & 1.50(-31) \\
				& 3  & 1.39(-02) & 1.95(-02) & 9.46(-03) & 8.61(-03) & 9.59(-04) & 5.43(-04) & 1.13(-11) & 3.10(-11) \\
				& 10 & 1.32(+00) & 3.79(-01) & 1.19(+00) & 3.48(-01) & 4.56(-01) & 1.65(-01) & 1.29(-03) & 1.52(-03) \\
				& 30 & 2.66(+01) & 1.15(+00) & 2.62(+01) & 1.14(+00) & 2.20(+01) & 9.95(-01) & 4.23(+00) & 2.55(-01) \\
				\hline
		\end{tabular}}
	\end{table*}
	
	
\end{document}